\documentclass[11pt,a4paper]{article}
\usepackage{amssymb} \usepackage{amsmath} \usepackage{graphicx}
\usepackage{slashed}
\usepackage{epsfig,latexsym,color,xcolor}
\usepackage{ulem}
\usepackage{amstext}    
\usepackage{array} 
\begin{document}
\def\Giulia{\bf\color{red}}
\def\bef{\begin{figure}}
\def\eef{\end{figure}}
\newcommand{\ans}{ansatz }
\newcommand{\be}[1]{\begin{equation}\label{#1}}
\newcommand{\beq}{\begin{equation}}
\newcommand{\ee}{\end{equation}}
\newcommand{\beqn}[1]{\begin{eqnarray}\label{#1}}
\newcommand{\eeqn}{\end{eqnarray}}
\newcommand{\bd}{\begin{displaymath}}
\newcommand{\ed}{\end{displaymath}}
\newcommand{\mat}[4]{\left(\begin{array}{cc}{#1}&{#2}\\{#3}&{#4}
\end{array}\right)}
\newcommand{\matr}[9]{\left(\begin{array}{ccc}{#1}&{#2}&{#3}\\
{#4}&{#5}&{#6}\\{#7}&{#8}&{#9}\end{array}\right)}
\newcommand{\matrr}[6]{\left(\begin{array}{cc}{#1}&{#2}\\
{#3}&{#4}\\{#5}&{#6}\end{array}\right)}
\newcommand{\cvb}[3]{#1^{#2}_{#3}}
\def\lsim{\raise0.3ex\hbox{$\;<$\kern-0.75em\raise-1.1ex
e\hbox{$\sim\;$}}}
\def\gsim{\raise0.3ex\hbox{$\;>$\kern-0.75em\raise-1.1ex
\hbox{$\sim\;$}}}
\def\abs#1{\left| #1\right|}
\def\simlt{\mathrel{\lower2.5pt\vbox{\lineskip=0pt\baselineskip=0pt
           \hbox{$<$}\hbox{$\sim$}}}}
\def\simgt{\mathrel{\lower2.5pt\vbox{\lineskip=0pt\baselineskip=0pt
           \hbox{$>$}\hbox{$\sim$}}}}
\def\unity{{\hbox{1\kern-.8mm l}}}
\newcommand{\eps}{\varepsilon}
\def\ep{\epsilon}
\def\ga{\gamma}
\def\Ga{\Gamma}
\def\om{\omega}
\def\omp{{\omega^\prime}}
\def\Om{\Omega}
\def\la{\lambda}
\def\La{\Lambda}
\def\al{\alpha}
\def\beq{\begin{equation}}
\def\eeq{\end{equation}}
\newcommand{\ov}{\overline}
\renewcommand{\to}{\rightarrow}
\renewcommand{\vec}[1]{\mathbf{#1}}
\newcommand{\vect}[1]{\mbox{\boldmath$#1$}}
\def\tm{{\widetilde{m}}}
\def\mcirc{{\stackrel{o}{m}}}
\newcommand{\Dm}{\Delta m}
\newcommand{\dm}{\varepsilon}
\newcommand{\tanb}{\tan\beta}
\newcommand{\nbar}{\tilde{n}}
\newcommand\PM[1]{\begin{pmatrix}#1\end{pmatrix}}
\newcommand{\up}{\uparrow}
\newcommand{\down}{\downarrow}
\newcommand{\refs}[2]{eqs.~(\ref{#1})-(\ref{#2})}
\def\omE{\omega_{\rm Ter}}
\newcommand{\eqn}[1]{eq.~(\ref{#1})}
%

\newcommand{\DSUSY}{{SUSY \hspace{-9.4pt} \slash}\;}
\newcommand{\DCP}{{CP \hspace{-7.4pt} \slash}\;}
\newcommand{\mc}{\mathcal}
\newcommand{\gr}{\mathbf}
\renewcommand{\to}{\rightarrow}
\newcommand{\gtc}{\mathfrak}
\newcommand{\wh}{\widehat}
\newcommand{\br}{\langle}
\newcommand{\kt}{\rangle}


\def\lsim{\mathrel{\mathop  {\hbox{\lower0.5ex\hbox{$\sim$}
\kern-0.8em\lower-0.7ex\hbox{$<$}}}}}
\def\gsim{\mathrel{\mathop  {\hbox{\lower0.5ex\hbox{$\sim$}
\kern-0.8em\lower-0.7ex\hbox{$>$}}}}}

\def\nn{\\  \nonumber}
\def\de{\partial}
\def\brf{{\mathbf f}}
\def\bbf{\bar{\bf f}}
\def\bF{{\bf F}}
\def\bbF{\bar{\bf F}}
\def\bA{{\mathbf A}}
\def\bB{{\mathbf B}}
\def\bG{{\mathbf G}}
\def\bI{{\mathbf I}}
\def\bM{{\mathbf M}}
\def\bY{{\mathbf Y}}
\def\bX{{\mathbf X}}
\def\bS{{\mathbf S}}
\def\bb{{\mathbf b}}
\def\bh{{\mathbf h}}
\def\bg{{\mathbf g}}
\def\bla{{\mathbf \la}}
\def\bmu{\mathbf m }
\def\by{{\mathbf y}}
\def\bmu{\mbox{\boldmath $\mu$} }
\def\bsig{\mbox{\boldmath $\sigma$} }
\def\bunity{{\mathbf 1}}
\def\cA{{\cal A}}
\def\cB{{\cal B}}
\def\cC{{\cal C}}
\def\cD{{\cal D}}
\def\cF{{\cal F}}
\def\cG{{\cal G}}
\def\cH{{\cal H}}
\def\cI{{\cal I}}
\def\cL{{\cal L}}
\def\cN{{\cal N}}
\def\cM{{\cal M}}
\def\cO{{\cal O}}
\def\cR{{\cal R}}
\def\cS{{\cal S}}
\def\cT{{\cal T}}
\def\eV{{\rm eV}}

\numberwithin{equation}{section}

\begin{flushright}
CERN-TH-2016-232\\
\end{flushright}
\vspace{6mm}

\large
 \begin{center}
 {\Large \bf Glimpses of  black hole formation/evaporation \\ in highly inelastic, ultra-planckian  string collisions}

 \end{center}

 \vspace{0.1cm}

 \vspace{0.1cm}
 \begin{center}
{\large Andrea Addazi}\footnote{E-mail: \,  andrea.addazi@infn.lngs.it} \\

{\it \it Dipartimento di Fisica,
 Universit\`a di L'Aquila, 67010 Coppito, AQ \\
LNGS, Laboratori Nazionali del Gran Sasso, 67010 Assergi AQ, Italy}
\end{center}

  \begin{center}
{\large Massimo Bianchi}\footnote{E-mail: \, massimo.bianchi@roma2.infn.it}
\\
{\it Dipartimento di Fisica, Universit\`a di Roma Tor Vergata, \\
I.N.F.N. Sezione di Roma Tor Vergata, \\
Via della Ricerca Scientifica, 1 00133 Roma, Italy}
\end{center}

  \begin{center}
{\large Gabriele Veneziano}\footnote{E-mail: \, gabriele.veneziano@cern.ch}
\\
{\it Coll\'ege de France, 11 place M. Berthelot, 75005 Paris, France  \\
Theory Division, CERN, CH-1211 Geneva 23, Switzerland,\\
 Dipartimento di Fisica, Universit\`a di Roma La Sapienza, 00185 Rome, Italy}
\end{center}

\vspace{1cm}
\begin{abstract}
\large
We revisit possible glimpses of black-hole formation by looking at  ultra-planckian string-string collisions at very high final-state multiplicity. 
We compare, in particular,   previous results using
the optical theorem, the resummation of ladder diagrams at arbitrary loop order, and the AGK cutting rules,  with the more recent study of $2 \rightarrow N$ scattering at $N \sim s  M_P^{-2} \gg 1$.
We argue that some apparent tension between the two approaches disappears once a reinterpretation of the latter's results  in terms of suitably defined infrared-safe cross sections is adopted.
Under that assumption, the typical final state produced in an ultra-planckian collision does indeed appear to share some  properties with those expected from the evaporation of a black hole of mass $\sqrt{s}$, although no sign of thermalization is seen to emerge at this level of approximation.

\end{abstract}
\newpage

\baselineskip = 20pt

\section{Introduction }

Ultra-planckian string-string collisions represent a perfect {\it gedanken experiment} where one can address fundamental issues on the merging of gravitational and quantum physics  within a consistent framework. One of the main aims of such a program is to understand whether and how quantum information is recovered in a process which, classically, would lead to  black-hole formation \cite{Eardley:2002re, Kohlprath:2002yh, Yoshino:2004mm, Giddings:2004xy} 
and, semi classically, to its Hawking evaporation \cite{Hawking:1974sw}. This program has been carried out since about thirty years along two different lines:
Gross and Mende (later joined by Ooguri)  \cite{Gross:1987kza,Gross:1987ar} computed the high-energy, fixed angle behavior of string scattering amplitudes at arbitrary genus. Since higher and higher genus contributions were found to be more and more important in that kinematic regime, they concluded that the string loop expansion diverges. The physical reason for such a result is clear: order by order the fixed-angle string scattering amplitude is exponentially suppressed, while physically it should be sizeable owing to Einstein's  gravitational deflection formula. Unfortunately, a Borel resummation of the divergent series \cite{Mende:1989wt}  can only be justified in a region of parameters where the  process is classically forbidden and, consequently, the cross section is still exponentially small.

A very different approach was taken by Amati, Ciafaloni and one of us 
\cite{Amati:1987wq,Amati:1987uf,Amati:1988tn,Amati:1990xe,Amati:1993tb}
 (hereafter referred to as ACV). One starts by working in energy or equivalently gravitational radius $$R_S = 2 G\sqrt{s}$$  and impact parameter $b$  space attempting an all-loop resummation. This is possible at arbitrarily high energy provided the impact parameter is also correspondingly high. In this region classical gravitational deflection as well as tidal effects \cite{Amati:1987uf, Giddings:2006vu} due to the string's finite size, are effectively recovered  within a unitary $S$-matrix framework provided one is far away from the expected gravitational-collapse region $b \sim R_S$ and as long as the tide-excited states are included in the Hilbert space (see \cite{D'Appollonio:2013hja} for a detailed study of that unitary $S$-matrix). 
The regime of classical gravitational collapse can be approached -- but unfortunately not (yet) entered
-- from two different directions in parameter space (see Fig. 1):
\begin{itemize}
\item By letting $b/R_S$ approach a critical value of ${\cal O}(1)$ while keeping both $b$ and $R_S$ much larger than the string length $\ell_s$. This turns out to be quite difficult, although some interesting progress has been made over the past ten years 
\cite{Amati:2007ak,Marchesini:2008yh,Veneziano:2008zb,Veneziano:2008xa,Ciafaloni:2008dg,Ciafaloni:2009in,Ciafaloni:2011de,Ciafaloni:2014esa}. We note, in particular, a recent result \cite{Gruzinov:2014moa},\cite{Ciafaloni:2015vsa},\cite{Ciafaloni:2015xsr} on the form of gravitational brems-strahlung in the regime of small deflection angles, suggesting the emergence of a typical energy scale for the emitted gravitons of order the Hawking temperature $$T \sim T_H = \hbar/R_S \quad .$$
\item By approaching the limit $R_S \rightarrow \ell_s$ from below. In this case life is easier since one can justify the validity of a (string corrected) leading eikonal approximation not suffering from the nasty classical corrections that make things complicated in the previous case.
Here one can make contact with the GMO regime (finding perfect agreement with their Borel resummation) but can also try to go further 
\cite{Amati:1987uf,Veneziano:2004er,Veneziano:2005du} towards the expected black-hole formation regime $b < \ell_s, R_S \to \ell_s$ \cite{Veneziano:1986zf,Letessier:1995ic,Susskind:1993ws,Halyo:1996vi,Halyo:1996xe,Horowitz:1997jc,Damour:1999aw}
while keeping some control over unitarity. It was found, in particular, that,  by taking into account the opening of new channels corresponding to the imaginary part of graviton exchange in string theory, it was possible to obtain a unitary $S$-matrix within a Hilbert space containing, besides the tide-excited states, also those responsible for the above mentioned imaginary part (see section 2.2 for details).
\end{itemize}

In an independent study
 Dvali, Gomez, Isermann, L\"ust, and Stieberger (DGILS henceforth) \cite{Dvali:2014ila} succeeded in carrying out a calculation of tree-level high-energy large multiplicity scattering amplitudes both in quantum field and in quantum string theory.
The claim is that their results support the validity of the idea, proposed by Dvali and Gomez \cite{Dvali:2010bf,Dvali:2011aa}, according to which BH's can be portrayed as Bose-Einstein condensates at criticality.

One can  identify some tension between the results in \cite{Amati:1987uf} and those in \cite{Dvali:2014ila} since in the former (ACV) approach loop corrections are crucial in  restoring unitarity through an interplay of real production and virtual corrections controlled by the AGK cutting rules \cite{Abramovsky:1973fm}, while in the latter (DGILS) virtual corrections are largely ignored.

The main purpose of this paper is to try and understand the origin of this tension and to offer a solution of it through a reinterpretation of the result of \cite{Dvali:2014ila}.

The plan of the paper is as follows.

In Section 2 we will briefly review the results of ACV for ultra-Planckian scattering in String Theory, first in the weak-gravity regime and then in the so-called string-gravity regime where a string-corrected eikonal approximation can be justified all the way up into the expected threshold for classical black-hole formation. 
In Section 3, we review the results  obtained in \cite{Dvali:2014ila} for the $2 \rightarrow N$ scattering in string and field theory at very high energy and multiplicity.
In Section 4, after  pointing out  a tension between the two approaches,
we discuss the effect of adding to the calculation in \cite{Dvali:2014ila} both real and virtual soft gravitons, following Weinberg's classic treatment.
In Section 5, we reinterpret the results of \cite{Dvali:2014ila} claiming that one can  eventually reconcile the two methods. 
In Section 6, we conclude and draw directions for further investigation. In Appendix A we review the AGK formalism \cite{Abramovsky:1973fm, Bartels:2005wa}, and 
in Appendix B we discuss more details on the Weinberg soft B-factor \cite{Weinberg:1965nx} . 

\begin{figure}[t]
\centerline{\includegraphics [height=9cm,width=1\columnwidth] {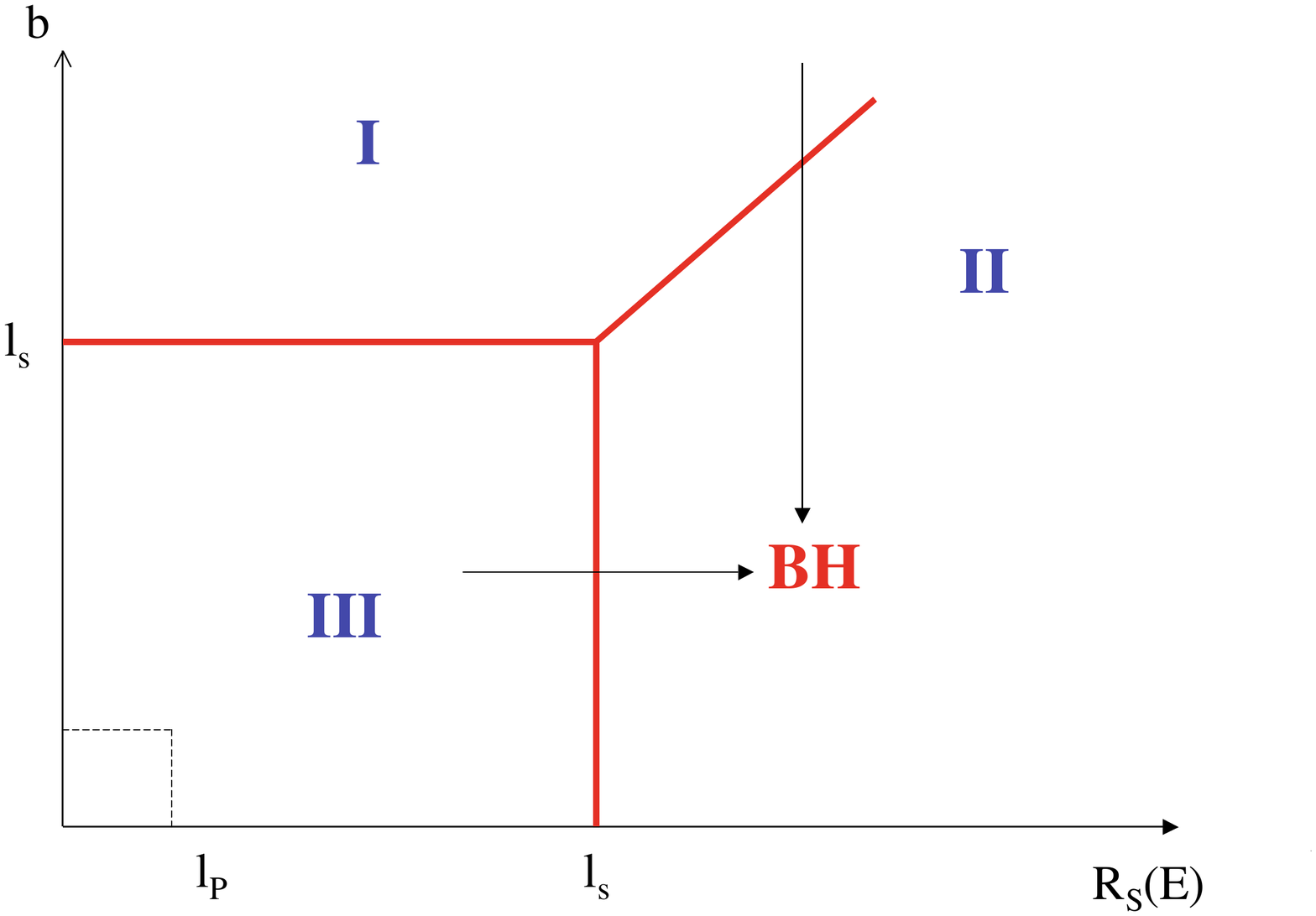}}
\vspace*{-1ex}
\caption{Rough phase diagram for transplanckian string-string collisions. In Region I and III calculations are essentially 
under control. They both border with region II where BH formation is expected on the ground of classical collapse criteria \cite{Eardley:2002re, Kohlprath:2002yh, Yoshino:2004mm, Giddings:2004xy}.}
\label{plot}   
\end{figure}

\section{The ACV approach: a  reminder}

For completeness we briefly review, in this section, some material that can be found in \cite{Amati:1987uf,Veneziano:2004er,Veneziano:2005du}.

\subsection{Different regimes in $b,R_S, \ell_s$ parameter space}
The main physical idea of the ACV approach is that, as long as ultra-planckian gravitational scattering is considered at sufficiently large impact parameter (in particular $b \gg R_S$), it is dominated by soft processes in which a very large number of nearly on-shell gravitons are exchanged between the two energetic particles. Such huge number of soft processes can build up a collision that we would otherwise call hard (at least in the language of QCD) because it  corresponds to fixed angle high energy scattering.

At very small (but finite) deflection angle the leading diagrams are simply $s$-channel ladder diagrams whose elementary rung is nothing but the tree-level graviton-graviton scattering amplitude given by\footnote{
Similar calculations of string scattering from a  stack of D-branes in the Regge regime 
 were used \cite{D'Appollonio:2015gpa} to show how 
Regge behavior saves string theory from possible  causality violations of the kind 
firstly noticed in Ref. \cite{Camanho:2014apa}.} 
\be{M1234}
\mathcal{M}_{0}(12\rightarrow 34)={\rm Tr}(h_{1}h_{4}){\rm Tr}(h_{2}h_{3}) \,{\cal A}_{0}(s,t)\, ,
\ee
where $h_i$ denote graviton polarisation tensors.  In the following we will set  $\alpha' =\ell_s^2 = 1/M_s^2 = 2$ (unless explicitly shown for clarity). For generic $s$ and $t$ ($s{+}t{+}u=0$) the amplitude in $D=4$ reads 
 $$\cM_{0}(12\rightarrow 34) = {2 g_s^2 \cR^4\over stu} 
 {\Gamma(1-s/2) \Gamma(1- t/2)\Gamma(1-u/2)\over \Gamma(1+s/2) \Gamma(1+t/2)\Gamma(1+ u/2)}\, , $$
 where $\cR^4$ denotes the contraction of 4 linearised Riemann tensors $\cR_{\mu\nu\rho\sigma} = k_{[\mu}h_{\nu][\rho} k_{\sigma]}$. 
 Taking the large $s$ limit and relying on Stirling formula the amplitude Reggeizes\footnote{The choice of the phase $(-)^{t/2} = e^{-i\frac{\pi}{2}t}$ is dictated by physical considerations. Since ${\rm Im}\cM$ must be positive at $t=-q^2 =0^-$, this is the correct choice.}  
\be{az}
{\cal A}_{0}(s,t)\simeq 4 g_s^{2}\frac{\Gamma(-t/2)}{\Gamma(1+t/2)}\left( \frac{s}{2}\right)^{\alpha(t)}e^{-i\frac{\pi}{2}t}\, ,
\ee
{\it i.e.} the scattering process proceeds through the exchange of the gravi-Reggeon trajectory with $\alpha(t)=2+\alpha' t/2 = 2+t$. Notice that while the real part of ${\cal A}_{0}(s,t)$ exposes the massless $t$-channel Coulomb pole, the imaginary part has no singularity for forward scattering. Moreover ${\rm Im} {\cal A}_{0}^{FS}(s) = \lim_{t\rightarrow 0} {\rm Im} {\cal A}(s,t)$ is related to the cross-section for production of (massive) string states at tree-level.

Clearly this amplitude is unfit to describe gravitational scattering. On one hand it is exponentially small at fixed (even small) angle while we expect a large cross section in that region from Einstein's gravitational deflection formula $\theta \sim R/b$. On the other hand, its Fourier transform (dominated by the fixed $t$ Regge region) gives a partial-wave amplitude ${\cal A}(J = b \sqrt{s}, s)$
that grows with energy thus violating unitarity bounds\footnote{N.B. Given the presence of massless particles one cannot use Froissart's bound. However, partial-wave unitarity still puts the constraint $|{\cal A}(J,s)| \le 1$}.

It proves convenient to resort to the eikonal approximation, whereby the dominant contribution to the $L$-loop amplitude reads$$
\mathcal{M}_{L}(s,t) \approx  (2\pi)^{D-2}\delta^{D-2}(q-\sum_iq_{i}) $$
\be{Ahst}
\times {{\rm Tr}(h_{1}h_{4}){\rm Tr}(h_{2}h_{3}) \over (L{+}1)!} \frac{i^L}{(2s)^L}\int \left[\prod_{i=1}^{L{+}1}{d^{D-2}q_{i}\over (2\pi)^{D-2}} {\cal A}_{0}(s,-q^2_{i}) \right] \mathcal{V}^2_{L{+}1}(q_i)\, ,   
\ee 
where $\mathcal{V}_{N}(q_i)$ denotes the $N$ gravi-reggeon vertex 
which in the limit $\alpha' q_{i}q_{j} \rightarrow 0$, 
reduces to 
$\mathcal{V}_{N}(q_i)=1+\mathcal{O}(\sum_{i<j}(\alpha'q_iq_j)^{2})$.

The amplitude $\mathcal{M}_{L}$ is a convolution in the $q$-space,
so that it can be factorized in the dual space of impact parameter $b$:
\be{As}
\frac{1}{s}\mathcal{M}_L(s,t)=4{\rm Tr}(h_{1}h_{4}){\rm Tr}(h_{2}h_{3})
\int d^{D-2}b\,e^{i qb}\tilde{\mathcal{A}}_L(s,b)\, ,
\ee
where
\be{aHsb}
\tilde{\mathcal{A}}_L(s,b)=\frac{(2i)^{L+1}}{(L+1)!}\langle 0| [\hat{\delta}(s,b,\hat{X}^{u},\hat{X}^{d})]^{L+1}|0 \rangle\, ,
\ee
with $\hat{\delta}$ the `eikonal' operator, related to the S-matrix by $$\hat{S} = 1+i\hat{T} = \exp 2i\hat{\delta} \quad .$$ As indicated, $\hat{\delta}$ is a functional of the closed string coordinates at equal time, and was found in \cite{Amati:1987wq} to take the highly suggestive form
\be{deltahat2}
\hat{\delta} (s,b;\hat{X}^{u},\hat{X}^{d}) =\int \frac{d^{D-2}q}{(2\pi)^{D-2}}\frac{{\cal A}_{0}(s,t)}{s}\int \frac{d\sigma_{u}d\sigma_{d}}{(2\pi)^{2}}:e^{iq(b+\hat{X}^{u}(\sigma_{u})-\hat{X}^{d}(\sigma_{d}))}:
\ee
$$=\int \frac{d\sigma_{u}d\sigma_{d}}{(2\pi)^{2}} :\tilde{\mathcal{A}}_{0}(s,b+\hat{X}^{u}(\sigma_{u})-\hat{X}^{d}(\sigma_{d})):\,\,,$$
corresponding to exchanging the graviton between one point on one string and one on the other.

When stringy effects are negligible, one can set $\hat{X}$ to zero and   
(\ref{deltahat2}) becomes an ordinary function, the eikonal phase, whose real part encodes elastic scattering  (with a physically irrelevant IR divergence in $d=D-4 = 0$), while the extra term associated with inelastic channels is finite. Setting
$$ Y= \log\alpha's \quad ,$$
and following \cite{Amati:1987wq} (briefly reviewed in Appendix A), one can perform  the integral by a saddle point method for $b^{2}>>\ell_{s}^{2}Y$, and obtain
\be{String}
\delta(b,s)=\left. \hat\delta(b,s)\right\vert_{\hat{X}=0}  \approx \left(\frac{b_{E}}{b}\right)^{d}+i\frac{G_{D}s}{\ell_{s}^{d}Y^{d/2+1}}e^{-b^{2}/Y\ell_{s}^{2}}\, ,
\ee 
where $G_{D}$ is the $D$-dimensional Newton's constant and 
$$b_{{E}}^{d}(s)={s \over 8\pi \Omega_{d} M_D^{d+2}} ={g_s^{2}s\over 8\pi \Omega_{d} M_s^{d+2}}\, ,$$
with $\Omega_{N}=2\pi^{N/2}\Gamma(N/2)$.

ACV distinguished several regimes in the $b, R_S$ plane (see Fig. 1):
\begin{itemize} 
\item The very large $b$ regime, $b>b_E$, Here the massless graviton pole dominates, though distorted by a Coulomb phase in $D=4$.
One recovers here, in a saddle-point approximation, the appropriate generalization of Einstein's deflection formula.
\item An intermediate  regime  $b_E> b> b_B, b_I$, where 
$$ b_I = \ell_{s}\sqrt{Y}=\ell_s \sqrt{\log\alpha's}$$
is the threshold for the opening of inelastic channels and 
$b_B =( b_E^d R^2)^{\frac{1}{d+2}}$
is the threshold for gravitational radiation {\it i.e.} the onset of $R_S/b$ corrections.
Here the Eikonal approximation applies. In the subregion $b_E> b_t > b> b_I$, where
$$b_{{t}}^{d+2}(s) ={s \over 8\pi \Omega_{d+2} g_{s}^{2}M_D^{d+4}} ={ g_{s}^{2} s \over 8\pi \Omega_{d+2} M_s^{d+4}}$$
determines the opening of inelastic channels, `tidal excitations' dominate
which are represented by excited string states.
The diffractive $b$-parameter emerges by
considering  second-order correction to the elastic channel part of the expression (\ref{String}) 
 by string finite size effects that modify the S-matrix.
\item The classical corrections regime $ b_B >  b > b_I$ where classical corrections and gravitational brems-strahlung kick in.

\item Inelastic regime $b_I> b> \ell_s > \ell_P$,  where inelastic channels of both classical and string absorption are opened. The relative importance of the two depends on whether $R_S$ is larger or smaller than $\ell_s$. For $R_S < \ell_s$ the situation is under control and string `softening' effects modify General Relativity in particular the deflection angle reaches a maximum around $b=b_I$ and then decreases again towards $b \approx \ell_s$ and then $b\approx R_S$. Since this is the regime of interest here it will be discussed separately and in greater detail in the next subsection.

\end{itemize}

\subsection{The string-gravity regime via the AGK cutting rules}
The  ``string-gravity" regime of ACV is defined (up to possible logs) by the inequality $\ell_s > b, R_S$. It is believed that in this regime the so-called classical corrections (that scale as $R_S^2/b^2$ in $D=4$) are tamed since they become, effectively, of order $R_S^2/\ell_s^2 \ll 1$. 
For this reason the string-gravity regime can be described in terms of the string-size-corrected leading eikonal approximation and, consequently, we do not expect to find here signatures of actual BH formation.
The string-corrected leading eikonal was already discussed in the previous subsection and leads to a unitary $S$-matrix which becomes highly inelastic at $b < b_t$ {\it i.e.} when tidal excitations of the incoming strings dominates. This phenomenon will persist in the string-gravity regime; however, a new source of inelasticity takes place on top of the one due to tidal excitation.

The origin of this new source of inelasticity can be easily ascribed to the fact that
in string theory the gravitons Reggeize, {\it i.e.} full Regge trajectories (starting from the graviton) are exchanged between the high-energy colliding particles (that we have taken to be massless gravitons) \cite{Amati:1987wq, Amati:1987uf}. We will be referring to the exchanged objects, therefore, as ``gravi-reggeons" (GR). That implies that the amplitude due a single GR exchange exhibits both a real and an imaginary part.
The former has the Coulomb pole (as in the QFT limit of ordinary graviton exchange) and, correspondingly, has a large-$b$ tail, while the latter is negligible at $b \gg \ell_s \sqrt{Y}$ but becomes relevant in the opposite regime 
$b \ll \ell_s \sqrt{Y}$ that includes the just defined string-gravity one. This imaginary part simply corresponds to the on-shell $s$-channel closed strings which are dual (in the old sense of DHS duality, after Dolen, Horn and Schmid) to the GR's. 
As in \cite{Veneziano:2004er,Veneziano:2005du} we will refer to such objects as ``cut gravi-reggeons" (CGR). Mathematically, this implies that the eikonal operator $\hat{\delta}$ of Eq. (\ref{deltahat2})
ceases to be hermitian. In order to restore formally unitarity one needs to introduce \cite{Veneziano:2004er,Veneziano:2005du} new creation and destruction operators $C$ and $C^\dagger$ for the on-shell states corresponding, in a broad sense, to a single CGR exchange\footnote{Obviously, in order to have full control of unitarity one should introduce separate, mutually commuting creation and destruction operators for each closed string contained, with a specific amplitude, in a single CGR. This remains, for the moment, an unfinished task.}. 

In \cite{Veneziano:2004er,Veneziano:2005du} it was pointed out that a formal way to recover a unitary $S$ matrix when $\hat{\delta}$ is not hermitian consists of the replacement:
\be{prescr}
\exp(2 i \hat{\delta}) \to \exp\left( i (\hat{\delta} + \hat{\delta}^{\dagger}) \right) \exp\left(i \sqrt{2i(\hat{\delta}^{\dagger}- \hat{\delta} )} (C + C^{\dagger})\right)\, .
\ee

Consider now the total cross section at some fixed impact parameter $b < l_s \sqrt{Y}$. By the optical theorem $\sigma_{\rm tot} (s) = \kappa {\rm Im}{{\cal A}}_{FS}(s)/s$, this will consist of a sum over all possible ways of ``cutting" the ladder diagrams that build up the leading eikonal approximation.

 Because of the above-mentioned nature of GRs, a ladder with $n$ GRs, being non-planar, can be cut along any number $n_c$ of GR with $ 0 \le n_c \le n$.
The problem of determining the relative weights for cutting a different number $n_c$ of GR is very similar to the one encountered in the sixties for an $n$-Pomerons exchange in hadronic physics and was nicely settled by the remarkably simple Abramovski-Gribov-Kancheli (AGK) cutting rules \cite{Abramovsky:1973fm}  (See Ref.\cite{Bartels:2005wa} for a useful review on these aspects in pQCD; a short reminder of which is given in Appendix A). These rules also follow directly from (\ref{prescr}) if one identifies $n_c$ with the number operator  $C^\dagger C$. We will not attempt to describe the operators $C$ and $ C^\dagger$ in details here but they may be related to the operators for higher spins originally defined by Weinberg 
  \cite{Weinberg:1964cn,Weinberg:1964ev,Weinberg:1969di}.

The AGK rules state that the $n_c$-CGR contribution to the full imaginary part of the elastic (fixed $b$)  $n$ gravi-Reggeon exchange amplitude\footnote{Following \cite{Veneziano:2005du}, we denote $\tilde\sigma = d\sigma/d^2b$ henceforth. The corresponding amplitude is denoted by ${\tilde{\cal A}}$.} by  is given by:  
\be{AAsigmamn}
\tilde\sigma^n_{n_c}=(-1)^{n-n_c}\frac{(4{\rm Im} \delta(s,b))^{n}}{n_c!(n-n_c)!}\quad {\rm for} \quad 1\leq n_c\leq n\, ,
\ee
and 
\be{AAsigma0n}
\tilde\sigma_{0}^{n}=(-1)^{n}\frac{(4{\rm Im}\delta(s,b))^{n}}{n!}+2 {\rm Im} {\tilde{\cal A}}_{n} \quad {\rm for} \quad n_c=0\, .
\ee
Accordingly, the sum over all contributions correctly reproduces the total imaginary part of the amplitude in agreement with the optical theorem. For partial-wave unitarity it is however more transparent to work with the full $S$ matrix without extracting the no-transition term. In that case one should directly check that $S(s,b)S^{\dagger}(s,b) = 1$. 

Let us check this constraint by considering the following
more detailed formulation of the AGK rules: the  $n$-GR-exchange contributions to $S S^{\dagger}$ can be split according to the number $n_c$ of CGR, the number $n_+$ of  GR in $S$ and the number $n_-$ of GR in $S^{\dagger}$ according to:
\be{AGKrule}
S S^{\dagger}_{(n)} =  \sum_{ n_+ + n_- + n_c = n} \frac{(2 i \delta)^{n_+} (-2 i \delta^{\dagger})^{n_-}( 4{\rm Im}\delta)^{n_c}}{n_c! n_+! n_-!} ~;~ n \ge 1\, .
\ee
If we now keep $n_c = N$ fixed and sum over $n_+, n_-$ we reproduce (\ref{AAsigmamn}). Furthermore, if, for fixed $N$, we sum over all values of $n_+, n_-$ we get:
\be{sigmaN}
\tilde\sigma_{N} = e^{-4{\rm Im}\delta} \frac{(4{\rm Im}\delta)^N}{N!} \Rightarrow \sum_N \tilde\sigma_{N}  = 1 ~~;~~ i.e. ~~ S S^{\dagger} = 1\, .
\ee
A simple way to understand (\ref{AGK}) is as follows: the exchange of $n$ identical bosons carries a $1/n!$ weight. Multiplying this by the number of ways we can choose, out of them, the three subsets $n_+, n_-, n_c$  gives the combinatorial factor $(n_c! n_+! n_-!)^{-1}$. Finally, each set gives the appropriate $S$-matrix element or its imaginary part\footnote{One may be worried about energy conservation in the AGK rule, since the total CM energy $\sqrt{s}$ should be shared among the $n_c$ CGR, while $[{\rm Im}]\delta (s)$ is computed for the total $s$. A related issue is the (in)distinguishability of the $n$ exchanged particles out of which only $n_c$ are cut. Quite remarkably these two issues compensate in such a way that the AGK rules turn out to have such a simple form as in Eq. (\ref{AGKrule}).}.

It proves convenient to construct a generating function for the cross sections
\be{generating}
\Sigma(z)=\sum_{n}^{\infty}[\tilde{\sigma}_{0}^{n}+\sum_{m=1}^{n}z^{m}\tilde{\sigma}_{m}^{n}]=e^{4(z-1){\rm Im}\delta}\, ,
\ee
 which  is related to the S-matrix introduced in  \cite{Veneziano:2005du} through  
\be{gfS}
\Sigma(z)=\langle in| S^{\dagger}:e^{(z-1)\hat{N}}:S|in \rangle~;~ \hat{N} = C^{\dagger} C \, . 
\ee
Comparing with (\ref{generating}), one gets 
\be{NCGR}
\Sigma(z) = 4(z-1){\rm Im}\delta \Rightarrow \langle N\rangle =4 Im \delta = 4\,\frac{G_{D}s}{\ell_{s}^{d}Y^{d/2+1}}e^{-b^{2}/Y\ell_{s}^{2}}\, .
\ee
This result may be interpreted as the formation of a coherent state of gravi-reggeons 
with Poisson, rather than thermal black-body, distribution

\be{sigmamn}
\frac{d \sigma(2\rightarrow N)}{d^2 b} = \frac{\langle N \rangle ^{N}}{N!}e^{-\langle N\rangle} \quad ;   \qquad b < \ell_s \sqrt{Y}\, .
\ee
An alternative proof of (\ref{NCGR}) is given in Appendix A. 
  One can also evaluate the average energy per cut gravi-reggeon and find 
\be{ECGR}
\langle E \rangle=\frac{\sqrt{s}}{\langle N \rangle}= \frac{M_{s}^{2}}{g_{s}^{2}\sqrt{s}} Y \sim \frac{\hbar}{R_S} Y\, .
\ee
Note that 
 for $d=0, D=4$ the energy has an average value that coincides (modulo a $\log\alpha's$ factor) with 
the  Hawking temperature $T_{H}= \hbar/R_{S}$ of a would-be black hole whose temperature exceeds the Hagedorn temperature of string theory. Of course this is not the correct interpretation of the result.

Rather, we can say that, as the expected threshold of BH production $R_S = \ell_s$ is approached from below, the individual CGR have still an invariant mass$^2$ parametrically larger (albeit just by a $\log \alpha' s$) of the string scale, justifying the use of Regge behavior.

On the other hand, as one crosses into the strong-gravity ($R_S > \ell_s, b$) region, the expected energy of the individual CGR falls below the string scale. That means, physically, that each one of them should give rise to a bunch of massless strings. Taking seriously the logarithms, one might imagine that each CGR gives rise to ${\cal O}(Y^2)$ quanta of energy ${\hbar}/{R_S}$ for a total multiplicity of order $Gs/\hbar$ {\it i.e.} of the entropy of a ``String-Hole", a BH lying just at the corresponding curve between strings and black holes \cite{Susskind:1993ws, Halyo:1996vi, Halyo:1996xe, Horowitz:1997jc, Damour:1999aw}.

On the other hand, this is precisely the point at which a description of the final state in terms of massless particles alone should become reliable. And this is also the lower end of the regime discussed by \cite{Dvali:2014ila}. For consistency we would like the calculation described in this subsection and the one of \cite{Dvali:2014ila}, described in the next section, to smoothly join one another  along the correspondence line.

\section{The  classicalization approach to high-energy, high-multiplicity gravitational scattering}

Recently Dvali, Gomez and collaborators \cite{Dvali:2010bf,Dvali:2011aa,Dvali:2012en,Dvali:2012rt,Dvali:2010jz,Dvali:2010ns} proposed a quantum mechanical  description of black holes as Bose-Einstein condensates of a large number of gravitons ($N\approx M_{BH}^2/M_{Pl}^2$) that, at a critical value $\alpha^{\rm crit}_{G}/N\approx 1$ of the effective gravitational coupling $\alpha_{G} \approx G s$,  behave as Bogoliubov modes and form a BH bound state. The mechanism termed `classicalization' provides a quantum $N$-picture of BH's\footnote{More generically, `classicalization' represents a mechanism that provides the unitarization of a UV incomplete theory by means of the resonant production of a non-perturbative classical solution. 
An example in the context of non-local quantum field theory was studied in Ref. \cite{Addazi:2015ppa}.
} that has been tested in connection with ultra-planckian scattering and formation of black holes as self-critical Bose-Einstein condensate 
of soft gravitons in \cite{Dvali:2014ila,Kuhnel:2014xga}.  

In particular, in \cite{Dvali:2014ila}, relying on KLT relations (after Kawai, Lewellen, and Tye \cite{Kawai:1985xq}) and the `scattering equations' \cite{Cachazo:2013gna}
in the Regge limit, DGILS tried to demonstrate that the perturbative exponential suppression factor $e^{-N}$ with the number $N$ of produced particles (gravitons)\footnote{To be precise,  in \cite{Dvali:2014ila} the number of produced gravitons is $N{-}2$. In order to adhere to the original DGILS paper, in this section, we will follow this convention. Clearly $N{-}2 \approx N$ for $N>>1$.} is exactly compensated by the BH entropy 
in a self-critical phase. Let us summarise their derivation in  \cite{Dvali:2014ila} and later on comment on the issues raised by their analysis. 

The starting point for computing tree-level graviton amplitudes with large multiplicity of the final states are the `scattering equations' \cite{Cachazo:2013gna} and the KLT relations \cite{Kawai:1985xq, BjerrumBohr:2010zp, Stieberger:2013wea, Stieberger:2014hba}, that relate closed string amplitudes on the sphere to `squares' of open string amplitudes on the disk. In the `field-theory' limit KLT can be used to relate non-planar graviton amplitudes to color-ordered gluon amplitudes. The relevant formula reads \cite{Stieberger:2013wea, Stieberger:2014hba}
\beqn{MFFT}
 && \mathcal{M}_{\rm grav}(1,\ldots,N) = (-1)^{N-3}\kappa^{N-2}  \nonumber \\
&& \times \sum_{\gamma, \sigma \in S_{N-3}} \mathcal{A}^L_{YM}(1,\sigma,N-1,N)\mathcal{S}_{_{KLT}}[\gamma|\sigma]\mathcal{A}^R_{YM}(1,\gamma,N-1,N) \; ,
\eeqn  
where $\kappa^2 = 8\pi G = \ell_P^2$,  $S_{N-3}$ is the group of permutations of $N{-}3$ objects, and the KLT kernel is given by  
\be{SS}
\mathcal{S}_{_{KLT}}[i_{1},...,i_{k}|j_{1},...,j_{k}]=\prod_{t=1}^{k}\left(s_{i_{t},P}+\sum_{q>t}^{k}\theta(i_{t},i_{q})s_{i_{t},i_{q}}\right)\; ,
\ee
with $s_{ij}= 2k_ik_j$, $P$ an arbitrary reference light-like momentum, 
while $\theta(i_{a},i_{b})=0$ for $i_{b}$ in $\{j_{1},...,j_{k}\}$ and $\theta=1$ otherwise. 

In $D=4$ it is convenient to switch to the helicity spinor formalism, whereby light-like momenta are expressed as $k_\mu{=}\bar{u}(k) \sigma_\mu u(k)$ in terms of commuting Weyl spinors of opposite chirality $u_\alpha(k)$ and $\bar{u}_{\dot\alpha}(k)$. The latter are often denoted by $|k\rangle$ and $|k]$, and satisfy $u^\alpha(k)u_\alpha(k') = \epsilon_{\alpha\beta} u^\alpha(k)u^\beta(k') = \langle k,k'\rangle ={-}\langle k',k\rangle$ 
as well as $\bar{u}_{\dot\alpha}(k)\bar{u}^{\dot\alpha}(k') = \epsilon^{\dot\alpha\dot\beta} u_{\dot\alpha}(k)u_{\dot\beta}(k')= [k,k'] ={-}[k',k]$. As a result Mandelstam invariants can be written as $s_{ij}= 2k_ik_j = \langle i,j\rangle [j,i]$, with $\langle i,j\rangle = \sqrt{2 k_ik_j} \exp({i\phi_{ij}})$ and  
$[j,i]= \sqrt{2k_ik_j} \exp({-i\phi_{ij}})$, for real momenta.

For a Maximally Helicity Violating  (MHV) configuration of the graviton polarizations $h^{\pm2}_i = a_{i,L}^{\pm1}\otimes a_{i,R}^{\pm1}$, the relevant color-ordered YM tree-level amplitudes, coded in Parke-Taylor formula \cite{Parke:1986gb}, read
\be{YM}
\mathcal{A}_{YM}(1^{+},...,i^{-},...,j^{-},...,N^{+})=\frac{\langle i,j \rangle^{4}}{\langle 1,2\rangle \langle 2, 3\rangle...\langle N{-}1, N\rangle \langle N, 1\rangle}\, ,
\ee
where $\pm$ denote the helicity of the gluons, all considered as incoming.
Furthermore, assuming a very peculiar kinematical regime for the scattering process 
$1,2\rightarrow 3,..,N$, {\it viz.}
$s_{12}=s$, $t_{i}=s_{i(1,2)}=-s/N$, $s_{ij}=s/N^{2}$
with $i,j=3,..., N$, and relying on the `scattering equations' \cite{Cachazo:2013gna} DGILS obtain $\mathcal{S}_{_{KLT}}\sim \left( {s}/{N^{2}}\right)^{N-1}$ and
\be{AYM}
\mathcal{A}_{YM}\sim s^{(2-N)/2}f(\phi) N^{N},\,\,\,\,{\rm for}\,\,\, i^-,j^-=1,2
\ee
$$\mathcal{A}_{YM}\sim s^{(2-N)/2}f(\phi)N^{N-2}, \,\,\,\,{\rm for} \,\,\,i^-=3,..,N,\,\,\,j^-=1,2$$
$$\mathcal{A}_{YM}\sim  s^{(2-N)/2}f(\phi)N^{N-4},\,\,\,\,{\rm for} \,\,\,i^-,j^-=3,...,N \, ,$$
where $f(\phi)$ is a (complicated) function of the phases $\phi = \{\phi_{ij}\}$ of the spinor bilinears. 
Substituting (\ref{AYM}) into the KLT formula (\ref{MFFT}), for MHV amplitudes
DGILS get \be{gravamp}\mathcal{M}_{\rm grav}(1^{+2},...,i^{-2},...,j^{-2},...,N^{+2})\sim\kappa^{N}C_N s \, , \ee
where $\pm 2$ denote the helicity of the gravitons, all considered as incoming. Depending on whether the two negative helicity gravitons be in the initial (1,2) or final (3,\ldots,N) state one has 
$C_N=(N+1)! N^{2}$ for $i^{-2},j^{-2}=1,2$, 
$C_N=(N+1)!N^{-2}$ for $i^{-2}=1,2$ and $j^{-2}=3,..,N$ or {\it vice versa}
and $C_N=(N+1)!N^{-6}$ for $i^{-2},j^{-2}=3,...,N$. Similar results are found by DGILS for non-MHV configurations in the chosen kinematical regime, that in a sense should dominate the integral over the final phase-space. Assuming that the sum over polarisations produce a factor $c_H^N$ with $c_H\sim {\cal O}(1)$, DGILS estimate the cross section for a fixed but large number $N$ of gravitons to be of the form
\be{cross}
\sigma(2\rightarrow N{-}2)\sim N!\left(\frac{\ell_{P}^{2}s}{N^{2}}\right)^N\, .
\ee
Eventually, assuming self-criticality  {\it i.e.} $\alpha_G = \ell_{P}^{2}s=N = \alpha^{\rm crit}_G>>1$, 
\be{cross}
\sigma(2\rightarrow N{-}2) \sim \frac{N! }{N^{N}}\sim e^{-N}\, .
\ee
This result is quite surprising for various reasons that we would like to analyze in some detail in the following. Later on we will compare the classicalization approach with the eikonal approach and propose a way to reconcile the two.

\begin{itemize}
\item
In \cite{Dvali:2014ila} only tree level amplitudes in a specific kinematical regime of the final gravitons (suggested by classicalization) is considered. Neither explicit integration over the final phase space nor explicit sum over all possible helicity configurations are performed for obvious technical difficulties. 

\item In \cite{Dvali:2014ila} higher orders in perturbation theory, corresponding to virtual graviton exchange or `soft' graviton radiation, are not considered 
and stability of the results with respect to quantum corrections is tacitly assumed. In particular, it is not clear whether an exclusive cross-section is computed, that would be infinite due to IR divergences, or some sort of inclusive cross-section.

\item In \cite{Dvali:2014ila}, the criticality condition $\alpha_{G} = G s = c N$ with a precise, fine-tuned proportionality constant $c$ is assumed and eventually used in the cross section. The exponential factor $e^{-N}$ results from such a precise fine tuning. This factor is crucial for their classicalization argument: indeed, they argue,
if a resonant production of (micro) Black Holes mediates this channel, the
$e^{-N}$ factor can be compensated by the BH entropy factor $e^{+S_{BH}}\sim e^{N}$.
Note that, by a different choice of $c$, one can turn the exponential suppression into an exponential growth, reproducing the behavior of the (convergent) sum over $N$. In that case adding an entropy factor overshoot unitarity.
\end{itemize}

\section{A tension between the two approaches}

Clearly, there appears to be some tension between the results summarized in sections 2 and 3 while one would like the two to join smoothly at the expected threshold for BH formation.

The most obvious difference is that in the ACV/AGK approach virtual corrections (corresponding to uncut GR) are essential to restore unitarity at fixed $b$, while no extra phase space or number of states factor is needed. By contrast, in the DGILS approach no virtual corrections are included and the S-matrix (integrated over impact parameter) satisfies unitarity bounds thanks to an extra exponential entropy factor.
In the following we shall investigate the nature of  soft radiative corrections to the process discussed in DGILS, while in  section 5 we will offer a new interpretation of the DGILS result that appears to resolve the above mentioned tension.

To this end, we will recall Weinberg's theorem for soft gravitons  \cite{Weinberg:1965nx},
apply it to the tree-level amplitude computed by DGILS in the classicalization regime, and  
see how gravi-strahlung and virtual gravitons may affect the behavior of the cross-section in different kinematical regions.

\subsection{Virtual soft gravitons}

\begin{figure}[t]
\centerline{\includegraphics [height=5cm,width=0.8\columnwidth] {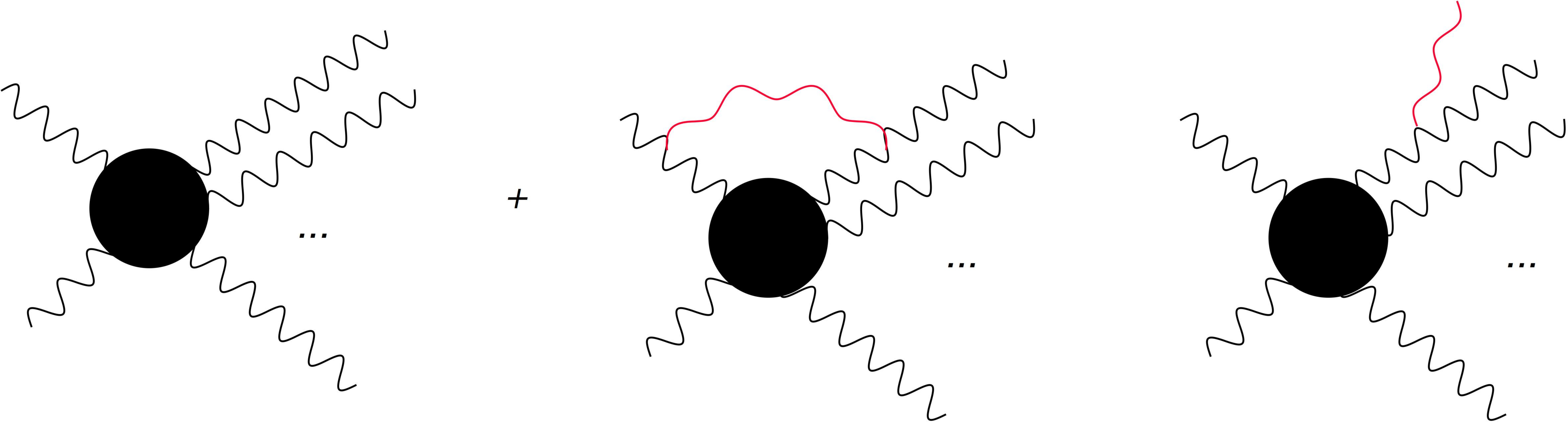}}
\vspace*{-1ex}
\caption{$2\rightarrow N$ graviton amplitudes with virtual soft gravitons corrections and emission of real soft gravitons from the external legs.}
\label{plot}   
\end{figure}

Let us start by considering the effect of adding a virtual soft graviton to graviton amplitudes. Although the final state of DGILS consists of gravitons of typical energy $M_p^2/\sqrt{s}$ we can always add a virtual graviton of even lower energy. In that case Weinberg has shown \cite{Weinberg:1965nx} that the single soft virtual correction amounts to multiplying the original amplitude by a factor:
\be{Alambda}
A_{1-IR-grav}=\frac{1}{2}B(p_i)A_{0}\quad ;\qquad B(p_i) = \sum_{i,j}\int_{\lambda}^{\Lambda} d^{4}q B(p_i, q)\, ; \nonumber
\ee

\be{Bq}
B(p_i,q) =  \frac{-8\pi i G}{(2\pi)^{4}[q^{2}-i\epsilon]}\sum_{i,j} \frac{\eta_{i}\eta_{j}\{(p_{i}{\cdot}p_{j})^{2}-\frac{1}{2}m_{i}^{2}m_{j}^{2}\}}{[p_{i}{\cdot}q-i\eta_{i}\epsilon][-p_{j}{\cdot}q-i\eta_{j}\epsilon]} \
; 
\ee
where $\lambda$ represents an IR cutoff, $\Lambda$ is an upper cutoff to be discussed later\footnote{Not to be confused with some UV cutoff of Quantum Gravity!}, while  $\eta=+1$ for outgoing particles and $\eta=-1$ for incoming particles. 

As shown in \cite{Weinberg:1965nx} this result can be generalized to the case of an arbitrary number of soft gravitons and the sum of all such contributions exponentiates so that:
\be{S}
S_{2\rightarrow M}=\mathcal{C}(\lambda,\Lambda) S_{2\rightarrow M}^{0}\, ,
\ee
where (suppressing the $p_i$ labels in $B(p_i,q) $)
\be{Norm}
\mathcal{C}(\lambda,\Lambda)= \sum_L \frac{1}{L!}\left[\frac{1}{2}\int_{\lambda}^{\Lambda}d^{4}q B(q)\right]^{L} \rightarrow {\rm exp}\left\{\frac{1}{2}\int_{\lambda}^{\Lambda}d^{4}q B(q) \right\}\, .
\ee
The corresponding correction to the rate reads
\be{rate}
|S_{2\rightarrow N}|^{2}=|S_{2\rightarrow N}^{0}|^{2}{\rm exp}\left\{{\rm Re} \int_{\lambda}^{\Lambda}d^{4}q B(q)\right\}\, ,
\ee
and depends only on the real part of the integral over the 4-momentum of the virtual graviton which
 only receives contribution from the imaginary part of the graviton propagator
$i\pi \delta(q^{2})$. One finally obtains\footnote{For completeness we give, in Appenxix B, a simple derivation of this result.}:
\beqn{onecon}
&&{\rm Re} \int_{\lambda}^{\Lambda}d^{4}q B(q)=-B_{0}\log(\Lambda/\lambda)~;~ B_{0} = \int d^{2}\Omega \frac{8\pi G}{2(2\pi)^{3}}\sum_{i,j}\frac{\eta_{i}\eta_{j}\left\{ (p_{i}{\cdot} p_{j})^{2}-\frac{1}{2}m_{i}^{2}m_{j}^{2}
\right\}}{[E_{i}-{\bf p}_{i}\cdot {{\vec n}}][E_{j}-{\bf p}_{j}\cdot {{\vec n}}]} \nonumber \\
&=& \frac{G}{2\pi}\sum_{i,j} \eta_{i}\eta_{j}m_{i}m_{j}\frac{1+\beta_{ij}^{2}}{\beta_{ij}(1-\beta_{ij}^{2})^{1/2}}\log \left(\frac{1+\beta_{ij}}{1-\beta_{ij}}\right)\quad ;\qquad  \beta_{ij} \equiv \left(1 - \frac{m_i^2 m_j^2}{(p_i \cdot p_j)^2} \right)^{1/2}\, .
\eeqn

It is straightforward  to take the massless limit of (\ref{onecon}) and to check that it is smooth and harmless (this is the well-known absence of collinear graviton divergences implied by the graviton's helicity $\pm2$). The result (that we have  found nowhere in the literature) is particularly simple:
\be{conv}
B_0  =  {2G\over\pi}  \left\{ {s} \log\frac{s}{\mu^2}+\sum_{{i}=1}^N \left[{t_{{i}1}} {\rm log}\frac{-t_{{i}1}}{\mu^2} + {t_{{i} 2}}{\log}\frac{-t_{{i}2}}{\mu^2}\right] +\sum_{{i}<{j}}^{1,N} 
s_{{i}{j}} {\rm log}\frac{s_{{i}{j}} }{\mu^2} \right\} \, ,
\ee
where $s = s_{12} = 2 p_1{\cdot}p_2$,  
$t_{{i},1/2} = - 2 p_{{i}}{\cdot}p_{1/2}$,  $s_{{i},{j}} = 2 p_{{i}}{\cdot}p_{{j}}$ and ${\mu^2}$ is an arbitrary mass scale that drops out since $$s_{12} = - \sum_i  t_{i1} = - \sum_j t_{2j} = \sum_{i<j} s_{ij}$$
thanks to momentum conservation. 
In the case of the 4-point amplitude, choosing ${\mu^2} = s = s_{12}$ for convenience, one simply finds
\be{stu}
B_{0} =  {4G\over \pi} \left( t {\rm log}\frac{-t}{s}+u {\rm log}\frac{-u}{s} \right) = - {4G s\over \pi} \left( \sin^2{\theta\over 2} {\rm log}\sin^2{\theta\over 2}+\cos^2{\theta\over 2}{\rm log}\cos^2{\theta\over 2} \right) \ge 0\, ,
\ee
with a maximum $B_{0} = + {4G s \log 2/ \pi}$ for $\theta = \pi/2$ and minima $B_{0} = 0$ for $\theta = 0, \pi$.
The situation is less clear for arbitrary but fixed $N$.  In the CM frame $p_1 = E(1,\vec{n})$, $p_2 = E(1,-\vec{n})$ and  $p_{{i}} = E_{{i}}(1,\vec{n}_{{i}})$ with $\vec{n}$ and $\vec{n}_{{i}}$ unit vectors. Momentum conservation yields $\sum_{{i}} E_{{i}}=2E$ and $\sum_{{i}} E_{{i}}\vec{n}_{{i}} =0$, setting $w_{{i}} = E_{{i}}/2E$  one has  
$$0 \le w_{{i}} \le 1/2 \quad , \quad \sum_{{i}} w_{{i}} =1 \quad , \quad \sum_{{i}} w_{{i}} \vec{n}_{{i}} =0 \, ,$$
and, after some algebra, one finds 
\beqn{B0trigo}
B_{0} &=&  {4G s \over \pi}  \left\{ {-} \sum_{{i}} w_{{i}} \left[ \sin^2\left({\theta_{{i}}\over 2}\right) \log \sin^2\left({\theta_{{i}}\over 2}\right)+ \cos^2\left({\theta_{{i}}\over 2}\right)\log \cos^2\left({\theta_{{i}}\over 2}\right) \right] \right. \nonumber \\
&+& \left. \sum_{{i},{j}} w_{{i}} w_{{j}}\sin^2\left({\theta_{{i},{j}}\over 2}\right)\log\sin^2\left({\theta_{{i},{j}}\over 2}\right) \right\} = B^+_{0} + B^-_{0}\, ,
\eeqn
where the single sum $B^+_{0}$ is positive and smaller than ${4G s \over \pi} \log 2$ while the second term
(double sum) $B^-_{0}$ is negative and larger than $- {4G s \over \pi} {1\over e}$. So {\it a priori} one may expect $\log 2 \ge  \pi B_{0}/4Gs \ge -1/e$. While the upper bound can be reached, the lower bound cannot, due to kinematical constraints. Later on we will show that $B_0$ is always non negative 
by relating it to the integral of the square modulus of the leading soft factor.

We have systematically studied the value of $B_0$ in (\ref{conv}) as a function of the kinematic configuration of the $N$-particle final state with the following conclusions (see also Appendix B):

\begin{itemize}
\item  $B_0$ is zero only in very special configurations. These correspond to the forward elastic amplitude (as always) and to final states in which the above two final gravitons are replaced by an arbitrary number of strictly collinear ones. As soon as one moves away from this configuration $B_0$ becomes positive.
\item When one goes a large amount away from the above special kinematical regions $B_0$ is typically of order $Gs$ times a function of ${\cal O}(1)$ of the angles which does not grow with the number of final particles.

\item Some examples:

\begin{figure}[t]
\centerline{\includegraphics [height=2cm,width=1\columnwidth] {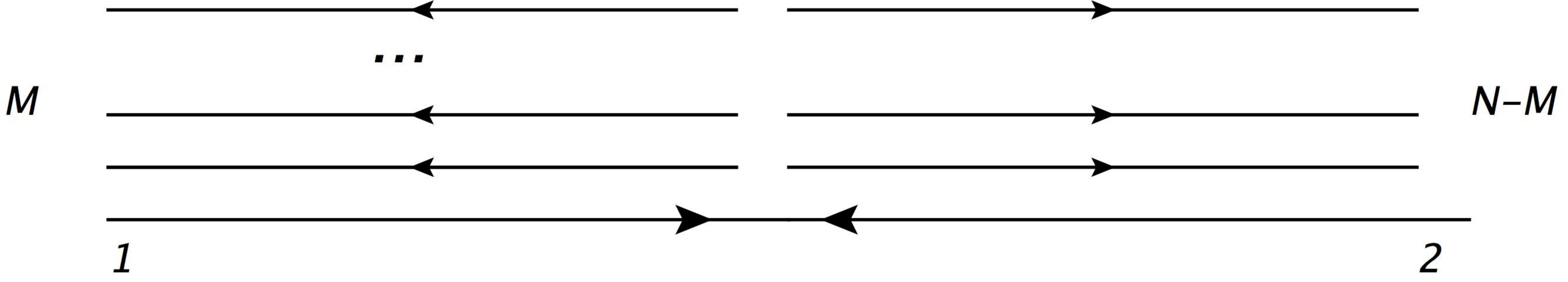}}
\vspace*{-1ex}
\caption{$2\rightarrow N$ scattering in which all the final gravitons are  collinear  to the initial ones. These are the only configurations with $B_0 =0$. }
\label{collinearprod}   
\end{figure}

\begin{figure}[t]
\centerline{\includegraphics [height=9cm,width=1\columnwidth] {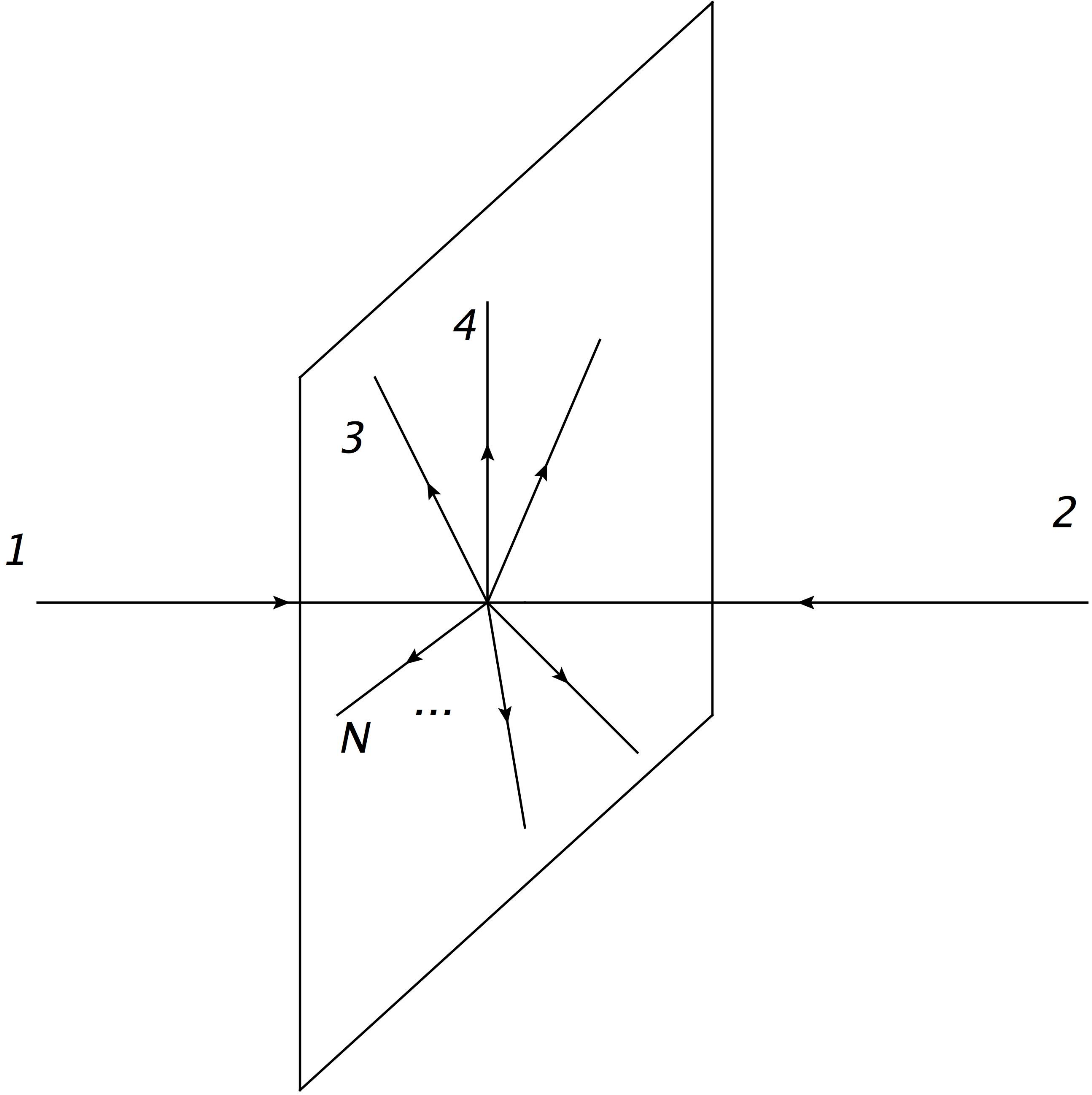}}
\vspace*{-1ex}
\caption{$2\rightarrow N$ scattering in which all the final gravitons are emitted in a plane orthogonal  to the initial momenta. Such configurations maximize $B_0$.}
\label{orthogonalprod}   
\end{figure}

{\bf Collinear} [Fig.\ref{collinearprod}]

For $ \vec{n}_{{i}} = \vec{n} \sigma_{{i}}=0$ with $\sigma_{{i}} = \pm 1$ such that $\sum_{{i}} w_{{i}}\sigma_{{i}}=0$, one has 
\beqn{B0collin}
B_{0} =  {2G s \over \pi}\left\{ \sum_{{i},{j}} w_{{i}} w_{{j}} 
{1{-}\sigma_{{i}} \sigma_{{j}} \over 2}  \log {1{-}\sigma_{{i}} \sigma_{{j}}\over 2} - \sum_{{i}} w_{{i}} \left[ {1{-}\sigma_{{i}}  \over 2} \log {1{-}\sigma_{{i}} \over 2} + {1{+}\sigma_{{i}} \over 2} \log {1{+}\sigma_{{i}} \over 2} \right] \right\} \, , \nonumber \\
\eeqn
which vanishes since ${1 - \sigma  \over 2} \log {1 -\sigma \over 2} = 0$ for $\sigma =\pm 1$.

{\bf Orthogonal } [Fig.\ref{orthogonalprod}]

For $ \vec{n}_{{i}}\vec{n} =0$ ($ \vec{n}_{{i}} \perp \vec{n}$), one has 
\beqn{B0perp}
B_{0} =  {2G s \over \pi}\left\{ \log 2 + \sum_{{i},{j}} w_{{i}} w_{{j}} {1{-}\vec{n}_{{i}}\vec{n}_{{j}} \over 2}  \log {1{-}\vec{n}_{{i}}\vec{n}_{{j}} \over 2}   \right\}  \ge {2G s \over \pi} \left[\log 2 - {1\over e} \right] \, ,
 \eeqn
since $x\log x \ge -1/e$ for $0\le x\le1$, with  $ x = (1 - \vec{n}_{{i}}\vec{n}_{{j}} )/ 2 = \sin^2(\theta_{{{i}}{{j}}}/2)$.

\end{itemize}

We thus conclude that, as a consequence of soft virtual graviton corrections, the $2 \rightarrow N$ cross section goes to zero (except in a zero-measure phase space region):
\be{corr1}
|S_{\lambda}|^{2}\rightarrow \left(\frac{\lambda}{\Lambda}\right)^{B_{0}}|S_{0}|^{2} \sim \left(\frac{\lambda \sqrt{s}}{M_P^2}\right)^{B_{0}}|S_{0}|^{2}\, .
\ee
We are of course very familiar with such a phenomenon that is counterbalanced by emission of soft radiation, {\it i.e.} bremsstrahlung.

Combining the contributions from virtual soft gravitons and real ones, to be discussed momentarily,  one eventually gets
\be{corr2}
|S_{\lambda}|^{2}\rightarrow \left(\frac{\Delta E}{\Lambda}\right)^{B_{0}}|S_{0}|^{2} \, ,
\ee
 which is nothing but the well-known cancellation between real and  virtual soft-graviton divergent contributions to the cross section.
 
\subsection{Real soft gravitons}

At this point one has to include IR divergences arising from emission of real soft gravitons. 
The real soft emission (gravi-strahlung) typically contributes a factor
$({\Delta E}/{\lambda})^{B_{0}}$, where $\Delta E$ is the maximal energy allowed in the soft radiation (see Sect. 5 for further details).
Once more, in order to trust our treatment of soft graviton emission we have to take $E \le M_P^2/\sqrt{s} \sim \Lambda$.

 The emission of soft gravitons at tree level is governed by the universal behaviour \cite{Cachazo:2014fwa, Bianchi:2014gla, Bern:2014vva}  
  \be{softoper}
 {\cal M}_{N+1}(p_i; q) \approx \sum_{i=1}^N \left[ {p_i h p_i \over \eta_i qp_i}  + {p_i h J_i q \over \eta_iqp_i} +
  {q J_i h J_i q \over \eta_i qp_i}\right] {\cal M}_{N}(p_i)\, ,
 \ee
 where $q^\mu$ and $h_{\mu\nu}$ denote the momentum and the polarisation of the soft graviton, while $J^{\mu\nu}_i = p^\mu_i \partial/\partial p^i_\nu - p^\nu_i \partial/\partial p^i_\mu + S^{\mu\nu}_i$ denote the angular momentum operator acting on the momentum and polarisation of the `hard' particles. Compared to YM and QED not only the dominant and sub-dominant terms but also the sub-sub-dominant term is universal \cite{Cachazo:2014fwa, Bianchi:2014gla, Bern:2014vva}. This holds true whenever gravitons couple as minimally as possible i.e. in the absence of $\phi R^2$ interactions involving `dilatons', while $R^3$ (non susy) or $R^4$ would not spoil universality \cite{Bianchi:2014gla, Bern:2014vva, Bianchi:2015lnw, Bianchi:2015yta, DiVecchia:2015oba, DiVecchia:2015jaq, Bianchi:2016viy, Bianchi:2016tju}.
 
 Let us consider first the dominant behaviour that is universal even beyond tree-level, {\it i.e.} at any order in perturbation theory in any consistent quantum theory of gravity such as String Theory, and compute the effect of adding a soft graviton to a process 
 \be{softadd}
\left\vert{\cal M}_{N+1}(p_i; q)\right\vert^2 = \int {d^3q\over 2|q| (2\pi)^3} \sum_{s=\pm 2} \left\vert\sum_{i=1}^N {p_i h_s p_i \over \eta_i qp_i}\right\vert^2 \left\vert{\cal M}_{N}(p_i)\right\vert^2\, .
\ee
The sum over polarisations/helicities $s=\pm 2$, with $h^{-s}_{\mu\nu} = (h^{s}_{\mu\nu})^* $, produces a transverse traceless bi-symmetric tensor
\be{tttensor}
\sum_{s=\pm 2} h^s_{\mu\nu} h^{-s}_{\rho\sigma}  = \Pi_{\mu\nu,\rho\sigma} = {1\over 2} 
\left( \Pi_{\mu\rho}\Pi_{\nu\sigma} + \Pi_{\mu\sigma}\Pi_{\nu\rho} - \Pi_{\mu\nu}\Pi_{\rho\sigma}\right)\, ,
\ee 
where $\Pi_{\mu\nu}= \eta_{\mu\nu} - q_\mu \bar{q}_\nu -q_\nu \bar{q}_\mu$ with $\bar{q}^2=0$ and $\bar{q} q = 1$. Luckily most of the terms are irrelevant thanks to momentum conservation $\sum_i \eta_i p_i = 0$ and to the mass-shell condition $p_i^2=0$.
In fact {\it lo and behold}
\be{sumij}\sum_{i,j} {p^\mu_i p^\nu_i \over \eta_i qp_i} \Pi_{\mu\nu,\rho\sigma} {p^\rho_j p^\sigma_j \over \eta_j qp_j} = 
\sum_{i,j}  {\eta_i \eta_j (p_i p_j)^2  \over qp_i qp_j } \, .
\ee
Integration over the soft light-like momentum $q= (|q|, \vec{q}) = |q|(1, \vec{n})$ produces
\be{Bagain}
8\pi G \int {d^3q\over 2|q| (2\pi)^3} \sum_{i,j}  {\eta_i \eta_j (p_i p_j)^2  \over qp_i qp_j } = {2G\over \pi} \log{\Lambda\over \lambda}  \sum_{i,j}  \eta_i \eta_j (p_i p_j) \log {p_i p_j\over \mu^2}\, ,
\ee
where the $\log\mu$, which is harmless thanks to momentum conservation, arises from a logarithmic divergence over the Feynman parameter $\alpha$ that can be regulated by giving a small (common) mass to the `hard' particles as shown in Appendix B.

It would be very interesting to study the sub- and sub-sub-leading corrections in the soft limit to the above result that are expected to be universal in gravity, at least at tree level.

  \section{A reinterpretation of the DGILS result and resolution of its tension with ACV/AGK}

The tree-level exclusive cross section $\sigma^{tree}(2 \rightarrow N)$ is, strictly speaking, infrared divergent (by taking, for instance, two of the final particles to be hard and the remaining $N-2$ to be arbitrarily soft, see Sect. 4). It is thus clear that the result of DGILS has to be reinterpreted.

 \begin{figure}[t]
\centerline{\includegraphics [height=6cm,width=0.7\columnwidth] {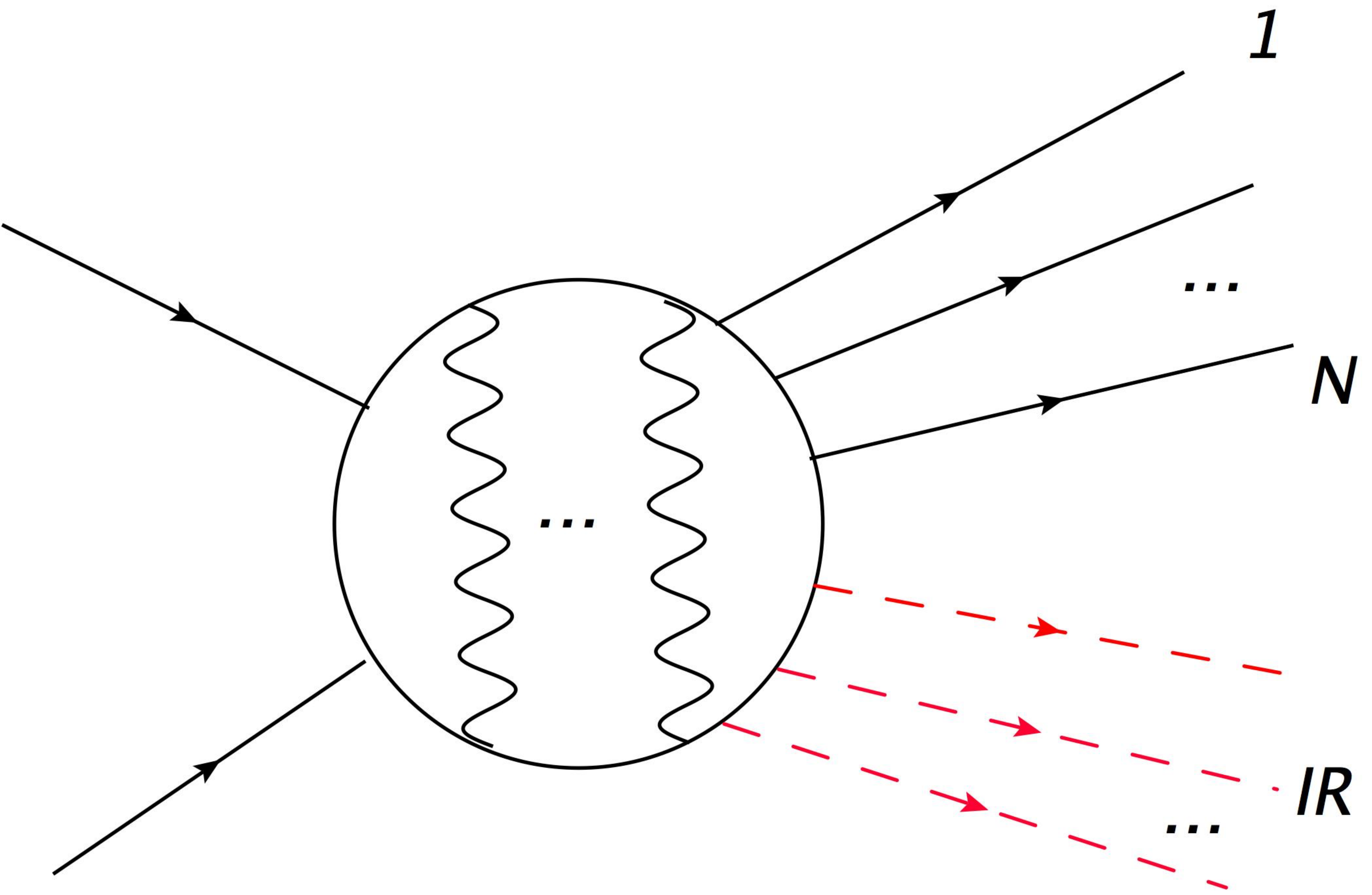}}
\vspace*{-1ex}
\caption{Representation of the $2\rightarrow N-{\rm jet}$ process. The $N$ hard ($E \ge \bar{E}$) gravitons are represented by solid outgoing lines, the IR gravitons ($E < \bar{E}$) by red dotted lines and the virtual gravitons by wiggly lines. }
\label{2toNvirtreal}   
\end{figure}

We shall follow, {\it mutatis mutandis}, a line suggested by the classic treatment 
\cite{Sterman:1977wj}
of jet cross sections in QCD. In that case one has to define observables which are insensitive
 to both infrared and collinear singularities. In the case of gravity the latter divergences are absent (that's why we could safely take the massless limit in Sect. 4) and therefore we shall pay attention to energies rather that to angular distributions. In the following integration over the angles has to be understood. An IR-safe quantity (carrying some analogy with the $N$-jet cross section in $e^+ e^- \to \rm{hadrons}$) is
 \be{defsigmaN}
\sigma\left(2 \rightarrow N (E_i \ge \bar{E})+ \rm{soft}(E_{\rm{soft}} \le \Delta)\right) \quad ; \qquad  \bar{E} N < \sqrt{s}\, ,
 \ee
in which the final state contains $N$ gravitons of c.m. energy greater than $\bar{E}$ and any number of soft gravitons of individual energy less than $\bar{E}$ and total energy less than $\Delta$ (called $\Delta E$ in Section 4), see Fig. \ref{2toNvirtreal}.
It is convenient, as in Sect. 2, to introduce a generating function(al) for such cross sections:
\beqn{defSigmazom}
&& \Sigma\left(  z(\omega), \bar{E}, \Delta \right) =   \\
 && \sum_{N,M} \int_{\bar{E}} d\omega_1 \dots d\omega_N  z(\omega_1) \dots  z(\omega_N)  \int_{\lambda}^{\bar{E}} d\epsilon_1 \dots d\epsilon_M 
\theta(\Delta - \sum_j \epsilon_j)
\frac{d^{N{+}M} \sigma}{ d\omega_1 \dots d\omega_N d\epsilon_1 \dots d\epsilon_M}\, , \nonumber
\eeqn
where we have denoted by $\omega_i, ~ i=1, \dots N$, the hard gravitons' energies and by $\epsilon_j, ~ j=1, \dots M$, those of the soft ones. The differential cross section carries a $\delta$-function for energy conservation $\delta(\sqrt{s} - \sum_i \omega_i - \sum_j \epsilon_j)$. Another $\delta$-function comes out if we consider the derivative of $\Sigma$ w.r.t. $\Delta$.

We shall be interested in studying $\Sigma$ together with some (functional) derivatives of it near $z(\omega) =1$. This will provide information on the ``total" cross-section as a function of $\bar{E}$ and $\Delta$ and on inclusive (one, two or more) hard graviton spectra. We shall carry out the first steps of this analysis below
after  introducing a simple ansatz for the differential cross section in (\ref{defSigmazom}). It will consist of three factors: the tree-level exclusive $2 \to N$ cross-section considered in \cite{Dvali:2014ila}, an additional factor accounting for the soft-graviton emission and, finally, a factor incorporating the effect of  virtual gravitons.

Concerning the tree-level differential cross section we shall reinterpret the result of \cite{Dvali:2014ila} as giving  the pre-factor that multiplies $d \log \omega_1 \dots d \log \omega_N$, i.e. we shall write: 
\be{prefactor}
\frac{d \sigma ^{\bf{tree}}}{ d\omega_1 \dots d\omega_N}  \sim N!  \left(c e^2 \frac{Gs}{N^2}\right)^N \frac{1}{\omega_1 \dots \omega_N} \sim \frac{1}{N!} (c ~Gs)^N  \frac{1}{\omega_1 \dots \omega_N}\, ,
\ee
where $c$ is some ${\cal O}(1)$ constant.

For the remaining two factors we simply use the results of Section 4 in order to write them in terms of the $B_0$ quantity introduced in (\ref{conv}). In the previous section we have seen that $B_0$
is of order $Gs$ unless the final particles are almost collinear with the initial ones in which case $B_0$ can be much smaller. In the following  we will approximate $B_0$ with some kind of average value  ${\tilde{c}}~ Gs$ where ${\tilde{c}}$ is another ${\cal O}(1)$ constant. This physically means that we are effectively excluding the large-$b$ (small deflection angle) regime  concentrating on the interesting one of small-$b$\footnote{Ideally one should instead project the results of \cite{Dvali:2014ila} and of Section 4 on partial waves (or fixed $b$) amplitudes, something non trivial and that we are deferring to further work.}. Using this approximation and the standard representation of the $\delta$-functions we  arrive at the following compact expression for ${\partial \Sigma}/{\partial \Delta}$:
\beqn{Sigmaz}
&& \frac{\partial \Sigma}{\partial \Delta}\left(  z(\omega), \bar{E}, \Delta \right) = \frac{1}{4 \pi^2} \int_{-\infty}^{+\infty}    d \sigma \int_{-\infty}^{+\infty} d \tau \exp\left( -i \sigma \sqrt{s}  -i  \tau \ \Delta \right) \nonumber 
\\
 &\times&\exp \left( c~Gs~ \int_{\bar{E}}^{\sqrt{s}} \frac{d \omega}{\omega} z(\omega)e^{i\omega \sigma} +  {\tilde{c}}Gs \int_{\lambda}^{\bar{E}} \frac{d \epsilon}{\epsilon} e^{i\epsilon ( \sigma + \tau)} - {\tilde{c}}Gs  \int_{\lambda}^{\Lambda} \frac{d \epsilon}{\epsilon} \right)\, .
\eeqn
Here we have also used some large-$N$ approximations that should not matter as far as we look at the neighborhood of $z(\omega) =1$. Finally, we have (re)introduced the cutoff parameter $\Lambda$ as an upper limit on the virtual gravitons' energy. We shall discuss its possible values below.

We will estimate (\ref{Sigmaz}) around $z=1$ through a saddle point approximation (in $\sigma$ and $\tau$) which should be reliable in our regime $Gs \gg 1$. However, before doing that, let us note that the single (hard) graviton distribution can be formally obtained from (\ref{Sigmaz}) through an appropriate functional derivative:
\be{singlegrd}
\frac{1}{\sigma} \frac{d \sigma}{d \omega} =  \left.\frac{\delta \log (\Sigma)}{\delta z(\omega)} \right\vert_{z=1} = \left<   c~Gs \frac{e^{i \sigma \omega}}{\omega}  \right>  \, ,
\ee
where the angled brackets mean the expectation value wrt the integrals over  $\sigma$ and $\tau$ appearing in (\ref{Sigmaz}). In the saddle point approximation this amounts to inserting the saddle-point value of $\sigma$ in (\ref{singlegrd}).

We now look for complex saddle points for the $\sigma$ and $\tau$ integrals of the form:
\be{defsaddle}
\sigma_{s} = i x /\sqrt{s}\quad , \quad \tau_{s} = i y /\Delta \, .
\ee
Imposing stationarity of the (large) phase w.r.t. $\sigma$ and $\tau$ we find:
\beqn{saddleeq}
x &=& c~Gs (e^{-\frac{\bar{E}}{\sqrt{s}} x} - e^{-x}) + {\tilde{c}}~Gs \frac{x}{(x+  y \sqrt{s}/\Delta)}\left(1- e^{-\frac{\bar{E}}{\sqrt{s}}(x+  y \sqrt{s}/\Delta)} \right) \nonumber \\
1 &=&  {\tilde{c}}~Gs  \frac{\sqrt{s}/\Delta}{(x+  y \sqrt{s}/\Delta~)}\left(1- e^{-\frac{\bar{E}}{\sqrt{s}}(x+  y \sqrt{s}/\Delta)} \right) \; ,
\eeqn
implying a condition involving just $x$
\be{saddleeq1}
x = c~Gs \left(e^{-\frac{\bar{E}}{\sqrt{s}} x} - e^{-x}\right) + \frac{\Delta}{\sqrt{s}} x ~~ \Rightarrow x   e^{\frac{\bar{E}}{\sqrt{s}}x} = c~Gs \, ,
\ee
where in the last equation we have used $\frac{\bar{E}}{\sqrt{s}}, \frac{\Delta}{\sqrt{s}} \ll 1$. Eq. (\ref{saddleeq1}) is solved by:
\be{saddlegen}
x = \frac{\sqrt{s}}{\bar{E}} W_0(\bar{E}/T_H) \, ,
\ee
where $W_0(z) = \sum_{n=1}^\infty (-n)^{n{-}1} z^n/n!$ is the (first branch of the) Lambert (or product-log) function.
 Inserting this result in the second of (\ref{saddleeq}) we find  that, quite generically, $y \sim Gs$. More precisely, under the assumption that $y \gg \frac{\Delta}{\sqrt{s}} x$, (\ref{saddleeq})  reduces to:
 \be{yeqsimpl}
 \hat{y} \equiv \frac{y}{Gs} = 1 - e^{- {\gamma} \hat{y} }\quad ;\quad {\gamma} \equiv \frac{\bar{E}}{T_H}\frac{\sqrt{s}}{\Delta}\, ,
 \ee
 whose solution is again given in terms of the Lambert function:
 \be{solyhat}
  \hat{y} = 1 + \frac{W_0(-{\gamma} e^{-{\gamma}})}{\gamma} \, ,
 \eeq
 showing that 
 $\hat{y} = {\cal O}(1)$ provided $({\gamma}- 1) = {\cal O}(1)$. Under this condition, $y \sim Gs$ and the inequality $y \gg \frac{\Delta}{\sqrt{s}} x$ can be easily checked for any value of $\bar{E}/T_H$.
 The above assumption on ${\gamma}$ looks very reasonable: actually, if ${\gamma} < 1$, eq. (\ref{yeqsimpl}) has no positive-$y$ solution.
 
 We also need to set a bound on $\Lambda$, the upper cutoff on virtual-graviton momenta. It seems obvious that such an upper bound should not be lower than the cutoff $\bar{\omega}$ on real (hard) gravitons.
 This can be easily estimated through eq. (\ref{singlegrd}):
 \be{cutoffhard}
 \bar{\omega} = \frac{\bar{E}}{W_0(\bar{E}/T_H)}\, .
 \ee
 Finally, let us write the result for $\Sigma(z=1)$ in a convenient form applicable to any value of $\bar{E}/T_H$):
 \be{Phi}
 \Phi \equiv \frac{\log \Sigma(z=1)}{Gs} = - \log\left(\frac{\Lambda \sqrt{s}}{T_H \Delta}\right) + E_1(W_0({\bar{E}}/{T_H})) \; ,
 \ee
where $E_1$ is the standard exponential integral function $E_1(x) = \int_x^{\infty} dy y^{-1} \exp{(-y)}$.
 
It is amusing to see the Hawking temperature $T_H$ emerge as a characteristic scale  distinguishing two cases:

1. $\bf{\bar E \ll T_H }$

In this case, using  $W_0(a) \sim  a$ for  $a \ll1 $ and taking $\Lambda \sim  \bar{\omega}$, we get: 
 \be{dsigma1}
 \Phi < - \log\left(\frac{\sqrt{s}}{\Delta}\right) - \log\left(\frac{\bar E }{T_H}\right) = - \log {\gamma} \le 0 \; ,
 \ee
 while, applying finally (\ref{singlegrd}), we obtain the suggestive result:
 
 \be{dsigmadom1}
 \frac{1}{\sigma} \frac{d \sigma}{d \omega} = \frac{Gs}{\omega} e^{- \frac{\omega}{ T_H}} \;,
 \ee
  exhibiting both a brems-strahlung behavior at small $\omega$ and a Boltzmann suppression at large $\omega$.

 2. $\bf{\bar E \gg T_H }$

In this case, using the large-argument limit of $W_0$: $W_0(a) \sim \log(a) - \log(\log(a))$ and again $\Lambda \sim  \bar{\omega}$, we get 
 \be{dsigma2}
 \Phi < - \log \left[\frac{\bar{E}}{T_H \log(\frac{\bar {E} }{T_H})} \frac{\sqrt{s}}{\Delta}\right] - \frac{T_H/\bar{E}}{\log(\bar E /T_H) }  <
 - \log \left[\frac{\bar{E}}{T_H \log(\frac{\bar {E} }{T_H})} \right] < 0 \; ,
  \ee
 while the analog of (\ref{dsigmadom1}) for the present case becomes:

 \be{dsigmadom2}
 \frac{1}{\sigma} \frac{d \sigma}{d \omega} = \frac{Gs}{\omega} e^{- \frac{\omega}{ \bar{E}}W_0({\bar{E}}/{T_H}) } \;,
 \ee
 joining smoothly with (\ref{dsigmadom1}) at $\bar{E} \sim T_H$.
 
  To summarize,
 under our mild assumption ${\gamma} -1 \ge {\cal O}(1)$, we have found two regimes as a function of ${\bar {E }}/{T_H}$:

  For $\frac{\bar {E }}{T_H} \gg 1$ the multi jet cross section is exponentially suppressed as $\exp(- \frac{G s}{\hbar} \log\frac{\bar{E}}{T_H} )$ and the hard-graviton spectrum is cut off at
   $\bar{\omega} \sim \frac{\bar{E}}{\log(\bar{E}/T_H)} > T_H$.
   
For $\frac{\bar {E}}{T_H} \ll 1$ the multi jet cross section can be ${\cal O}(1)$ and the hard-graviton cutoff is  $T_H$ independently of $\bar{E}$. 
To avoid an exponential suppression we need to take $\Lambda \sim T_H$ and ${\gamma} -1 =  {\cal O}(1)$ which looks physically possible. In particular, at $\bar{E} = T_H$ the fraction of energy in quanta below $T_H$ should be of ${\cal O}(1)$ which is also the case for black-hole evaporation.
Finally, the cutoff energy for the soft gravitons is $\frac{\Delta}{Gs}$. 
 
A final remark concerns our choice for $\Lambda$ and the virtual corrections attached to the incoming gravitons. For these corrections our estimate  $\Lambda \sim \bar{\omega}$ should be revised since the soft photon approximation can be now justified up to a scale smaller than but of order $\sqrt{s}$.
 These extra virtual contributions should have their own counterpart in real emission/absorption from the energetic legs. This is again what one should expect in the process of black-hole formation. In fact even before the critical impact parameter for BH formation is reached gravitational brems-strahlung takes place.
 While in the regime of small deflection angles this results in a small loss of energy (see \cite{Gruzinov:2014moa},\cite{Ciafaloni:2015vsa},\cite{Ciafaloni:2015xsr} for recent classical and quantum approaches to this problem) when the gravitational collapse regime is approached only a finite fraction of the incoming energy goes (at least classically) into forming a black hole.

\section{Summary and outlook}

A high-energy two-body collision usually results in the production of many lower-energy quanta. This is a common phenomenon shared by essentially any realistic $4-D$  interacting theory.
A well known and much studied case is the one of strong interactions in which  jets of hadrons are produced when the underlying parton shower (consisting of many final partons) hadronizes.
For low momentum transfer  processes the hadronic multiplicity typically grows logarithmically with energy (with the final particles filling up uniformly a rapidity plateau), while harder processes (such as $e^+ e^- \to {\rm hadrons}$) lead to multiplicities that typically grow faster than a power of $\log s$ (but slower than an exponential) of $\log s$ (e.g. like $\exp(\sqrt{\log s})$).

The fact that the gravitational coupling $G_N$ is dimensional and that the effective high-energy coupling is $\alpha_G \equiv \frac{G_N s}{\hbar}$ suggests that multiplicities in the unltraplanckian gravitational collisions should grow like a power of the center-of-mass energy $\sqrt{s}$. In particular, if $\langle n \rangle \sim \alpha_G$ we are immediately led --by energy conservation-- to the simple, yet startling conclusion that a {\it transplanckian} energy collision produces {\it subplanckian} final quanta of average energy $\langle E \rangle \sim \frac{\sqrt{s}}{\alpha_G } \sim \frac{\hbar}{R_S } \sim T_H$ with $T_H$ the Hawking temperature of a black hole of mass $\sqrt{s}$.

The study of ultraplanckian-energy collisions has confirmed to a large extent the above picture through what has been termed as ``fractionation". The first example \cite{Amati:1987wq,Amati:1987uf} is what we may call $t$-channel fractionation, i.e. the phenomenon by which a large momentum transfer is shared among  $n  \sim \alpha_G$ exchanged gravitons. This is the reason why a superficially hard process (fixed angle gravitational scattering at arbitrarily high energy) is actually controlled by large-distance physics with each exchanged graviton sitting very close to its mass-shell.

Within string theory $t$-channel fractionation has an $s$-channel analog  \cite{Amati:1987uf,Veneziano:2004er,Veneziano:2005du}. This is because, in string theory, graviton exchange is actually ``gravi-reggeon" exchange and carries an imaginary part related to the possibility of ``cutting" the graviton to expose the $s$-channel intermediate states dual to it. As a result, in the so-called string-gravity regime,
$t$-channel fractionation becomes, almost trivially, $s$-channel fractionation: the number of cut-gravi-reggeons grows with energy like $\alpha_G$ implying a softer and softer final state as one increases further and further $\sqrt{s}$. Unfortunately, this regime is under control only below a certain threshold energy corresponding to a Schwarzschild radius $R_S$ of order the string length $l_s$, i.e. to a would be Hawking temperature exceeding the Hagedorn temperature of string theory. And indeed we do not expect black-hole formation below such threshold.

In an independent development DGILS \cite{Dvali:2014ila} have addressed the problem of $s$-channel fractionation (called classicalization in their context) directly in quantum field theory i.e. without the use of string theory's duality. Evidence for fractionation would lend support to a previous proposal \cite{Dvali:2010bf,Dvali:2011aa} of black holes as a multi graviton state near a quantum phase transition. And, indeed, the claim in \cite{Dvali:2014ila} is that the $2 \to N$ cross section at $\alpha_G \gg 1$ is dominated by final states containing ${\cal O}(\alpha_G)$ quanta of energy ${\cal O}(T_H)$.

Taken at face value this result looks perfectly in line with the one of \cite{Amati:1987uf,Veneziano:2004er,Veneziano:2005du}, actually as a smooth extension of the latter in the theoretically unaccessible region above threshold. Although one result relies on string theory while the the other does not it is conceivable that, above the above mentioned threshold energy, the final state will consist of just massless strings for which a 
QFT approach is already sufficient.
However, at a closer scrutiny, some tension appears between the claims made within the two studies. While in \cite{Amati:1987uf,Veneziano:2004er,Veneziano:2005du} loop corrections to the $2 \to N$ process (corresponding to the possibility of cutting only a subset of the exchanged gravi-reggeons following the AGK rules) are crucial for restoring unitarity (or even just unitarity bounds),  in  \cite{Dvali:2014ila} one is only considering the $2 \to N$ process at tree level.

In this paper we have tried to resolve this tension by first considering soft real and virtual corrections to the process considered in  \cite{Dvali:2014ila} and by then showing that such corrections are in principle large in spite of the fact that true IR divergences cancel by the usual Bloch-Nordsiek mechanism. We have then given a reinterpretation of the claim in  \cite{Dvali:2014ila} through the introduction of some sort of ``gravitational jet cross sections", quantities that, like QCD jets, should be perturbatively calculable. They are characterized by a lower cutoff $\bar{E}$ on the energy of each jet-graviton and by an upper cutoff $\Delta$ on the total energy carried by all gravitons softer than $\bar{E}$.

So far we have only been able to estimate these jet cross section qualitatively. i.e. without control over ${\cal O}(1)$ parameters. This, however, is sufficient to support  the conclusion that, unlike QCD jets (that tend to be few and hard because of asymptotic freedom) gravitational jets tend to be many and soft, where {\it many} here means indeed ${\cal O}(\alpha_G)$ and soft means with energy of order $T_H$.
In other words we have found that, while the jet  cross section for $\bar{E} >> T_H$ is necessarily exponentially suppressed, the one with $\bar{E} \le T_H$ {\it can} be large enough to saturate unitarity. Furthermore, we have been able to study the one-jet inclusive cross section as function of its energy. We found that, in the relevant (latter) case such a distribution is bremsstrahlung-like ${dN}/{d\omega} \sim \omega^{-1}$ up to $T_H$ and Boltzmann-like suppressed (${dN}/{d\omega} \sim \exp(- \omega T_H^{-1})$) above. This falls short of agreeing with a Bose-Einstein thermal spectrum, for which presumably further re-interactions of the final gravitons have to be taken into account.

We should warn the reader that what was done here should only be considered as a first heuristic step into the problem of construction a unitary gravitational $S$-matrix sharing some properties with those of the semiclassical original analysis by Hawking. One problem that we left unanswered is that of projecting our results on individual $s$-channel partial waves (equivalently on a given impact parameter for large angular momentum), as done in  \cite{Amati:1987uf,Veneziano:2004er,Veneziano:2005du} but not in \cite{Dvali:2014ila}. Such an analysis should allow to see how the final state changes progressively from one typical of a scattering process to one resembling the evaporation of a black hole.
Here, in Section 5, we have considered directly the cross section integrated over impact parameter and it is only at the step where  we take the $B_0$ factor to be independent of the kinematics and of ${\cal O}(G s)$ that we are implicitly excluding peripheral processes (small deflection angles) concentrating our attention on the small-$b$ region.

We hope that the positive indications reached in this paper will motivate further work in this challenging --but hopefully highly rewarding-- line of research.
 
\vspace{1cm} 

{\large \bf Acknowledgments} 
\vspace{4mm}

We would like to thank M.~Bochicchio, M.~Ciafaloni, P.~Di~Vecchia, G.~Dvali, C.~Gomez, O.~Kancheli, F.~K\"uhnel, D.~L\"ust, J.~F.~Morales,  L.~Pieri, M.~Porrati,
S.~Stieberger, B.~Sundborg and C.-K.~Wen 
 for stimulating discussions and comments on a preliminary version of the manuscript. 
 A.~A would like to thank Fudan University for hospitality during the preparation of this paper. 
Work of A.~A. was supported in part by the MIUR research grant Theoretical Astroparticle Physics PRIN 2012CPPYP7 and by SdC Progetto speciale Multiasse La Societ\`a della Conoscenza in Abruzzo PO FSE Abruzzo 2007-2013.

\appendix

\section{A brief review of the AGK rules }

In a classic paper \cite{Abramovsky:1973fm}, Abramovski, Gribov and Kancheli (AGK) derived a set of ``cutting rules" allowing to relate the relative contribution of the ($t$-channel) exchange of $n$ Reggeons to different $s$-channel intermediate states. As a result, one can compute how the total cross-section, related by the optical theorem to the discontinuity wrt energy of the forward elastic amplitude, is shared among different final states.  The original derivation, reviewed in \cite{Bartels:2005wa}, was in the context of hadron scattering and the Pomeron. Here, following closely \cite{Bartels:2005wa},  we argue that it applies to ultra high energy gravitational scattering (within a string-theory context where the graviton belongs to a Regge trajectory) {\it mutatis mutandis}. 

Let us start by considering the Sommerfeld-Watson representation of the scattering amplitude 
for a binary $a+b\rightarrow a+b$ process \be{TAAB}
\mathcal{A}^{ab}(s,t)=\int \frac{d\omega}{2i}\zeta(\omega)s^{1+ \omega}\mathcal{F}^{ab}(\omega,t)\,\,\,,
\ee

where  $\omega = J{-}1$ is the conserved quantity in complex-angular-momentum theory and the signature factor $\zeta(\omega)$ reads 
\be{SWT}
\zeta(\omega)=\frac{\tau-e^{-i\pi \omega}}{\sin\pi \omega} = i + \frac{\tau-\cos\pi \omega}{\sin\pi \omega}\,\,\,,
\ee

with $\tau=\pm 1$ representing the signature. The integration contour in (\ref{TAAB}) is to the right of the singularities of $\mathcal{F}^{ab}$ (but to the left of those at the non-negative integer in $\zeta$).
Hereafter we shall only be interested in the case $\tau = +1$ for which, more simply:
\be{SWT+}
\zeta(\omega)= i + \tan \left(\frac{\pi}{2} \omega\right)\,\,\,.
\ee

The  amplitude $\mathcal{F}^{ab}(\omega,t)$ has poles and cuts
in the complex  ($t$-channel angular momentum) $\omega$-plane. 
In particular,  a non-planar multiple gravi-Reggeon exchange produces
branch cuts, just like ordinary particles do on the complex-energy plane.
In analogy with the latter case, the discontinuity of $\mathcal{F}^{ab}(\omega,t)$ across an $n$-Reggeon cut can be expressed as (see Fig. 5)
\be{expr}
{\rm disc}_{\omega}^{(n)}\mathcal{F}^{ab}(\omega,t)=2\pi i \int \frac{d\Omega_{n}}{n!}\Gamma_{\{\beta_{j}\}}\mathcal{A}_{n}^{a}(\{{\bf k}_{j},\omega\})\mathcal{A}_{n}^{b}(\{{\bf k}_{j},\omega\})\delta(\omega-\sum_{j}\beta_{j})\,\,\,,
\ee
where, setting ${\bf q}^{2}=-t$, the `transverse' $n$-particle phase space reads 
\be{dOmega}
d\Omega_{n}=(2\pi)^{2}\delta^{2}({\bf q}-\sum_{j=1}^n{\bf k}_{j})\prod_{j=1}^{n}\frac{d^{2}{\bf k}_{j}}{(2\pi)^{2}}\,\,\,,
\ee
with ${\bf k}_{j}$, $j=1,\ldots n$, denoting the transverse momentum of the $j$-th cut gravi-Reggeon. 
Furthermore, the product of all signature factors gives \cite{Bartels:2005wa}:
\be{gammabep}
\Gamma_{\{ \beta_{j}\}}=(-1)^{n-1}\frac{\cos[\frac{\pi}{2}\sum_{i}\beta_{i}]}{\prod_{i}\cos[\frac{\pi}{2}\beta_{i}] }\,\,\,,
\ee

with $\beta_{j}(-{\bf k}_{j}^{2}) = \alpha(-{\bf k}_{j}^{2}) - 1$ and $ \alpha(t_j= -{\bf k}_{j}^{2})$ the jth gravi-reggeon trajectory.

The vertex functions $\mathcal{A}^{a,b}_n$ represent the coupling of the external $a,b$ particles to $n$-Reggeons and depend on the process and the theory under consideration.

\begin{figure}[t]
\centerline{ \includegraphics [height=3cm,width=0.5 \columnwidth]{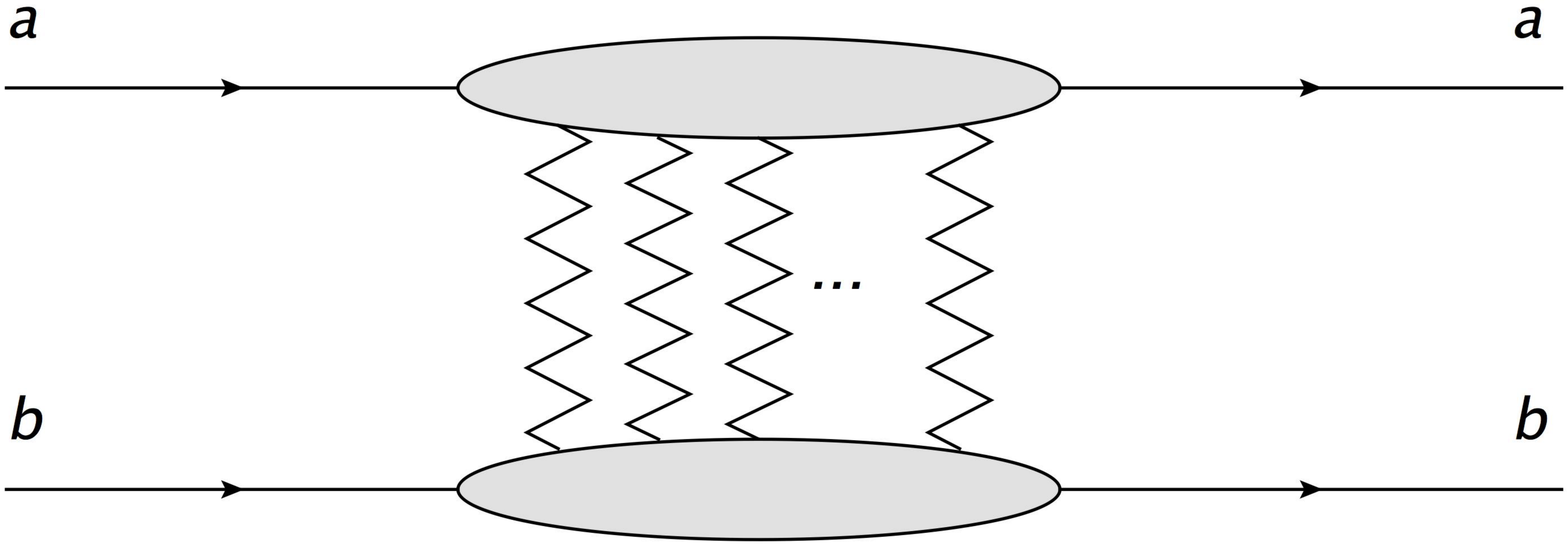}}
\vspace*{-1ex}
\caption{The $2\rightarrow 2$ amplitude with n-gravi-reggeons exchanged is parted in two $\mathcal{A}^{a,b}_{n}$ subamplitudes, diagrammatically schematizing
the Integral (\ref{expr}). 
}
\label{plot}   
\end{figure}
As a result, the contribution of $n$  gravi-Reggeons to the amplitude is given by 
\be{obt}
\mathcal{A}^{ab}_{n}(s, t) =\int \frac{d\omega}{2i}\zeta(\omega)s^{1+\omega}{\rm disc}_{\omega}^{(n)}\mathcal{F}^{ab}(\omega,t)
=\pi \int \frac{d\Omega_{n}}{n!}\Gamma_{\{\beta_{i}\}}\zeta(\sum_i\beta_{i})s^{1 + \sum \beta_{i}}\mathcal{A}_{n}^{a}\mathcal{A}_{n}^{b}\,\,\,.\ee

Let us note, at this point, that the combination $\Gamma_{\{\beta_{i}\}}\zeta(\sum_i\beta_{i})$ appearing in (\ref{obt}) takes the simple factorized form
\be{fact}
\Gamma_{\{\beta_{j}\}}\zeta(\sum_j\beta_{j} )= -i \prod_{j=1}^n \frac{- e^{-i \frac{\pi}{2} \beta_j}}{\cos(\frac{\pi}{2} \beta_j)} =  -i \prod_{j=1}^n (-1 + i \tan(\frac{\pi}{2} \beta_j))\,\,\,.
\ee

In the eikonal approximation, the vertex functions $\mathcal{A}^{a,b}_n$ factorize, i.e. $\mathcal{A}^{a,b}_n = (\mathcal{A}^{a,b}_1)^n$ and the expression (\ref{obt}) simplifies further if one goes over to impact parameter space by the Fourier transform:
\be{FT}
\mathcal{A}^{ab}_{n}(s, b ) = \int \frac{d^{2}{\bf q}}{(2 \pi)^2}e^{-i{\bf b}\cdot {\bf q}} \frac1s \mathcal{A}^{ab}_{n}(s, t= - q^2) \,\,\,.
\ee
We easily obtain:
\be{FTexp}
i \mathcal{A}^{ab}_{n}(s, b ) = \frac{1}{n!} \left(i \int \frac{d^{2}{\bf k}}{(2\pi)^{2}}  \mathcal{A}^{a}_1\mathcal{A}^{b}_1 s^{\beta(-k^2)} (i +  \tan(\frac{\pi}{2} \beta(-k^2)) \right)^n\,\,\,.
\ee

Summing finally over $n$ and adding 1 to go over to the $S$-matrix we recover the well-known eikonal exponentiation:
\beqn{eikonalb}
&& S^{ab} (s, b) = 1 + i \sum_{n=1}^{\infty} \mathcal{A}^{ab}_{n}(s, b )  \nonumber  \\
&=& \exp\left\{  i \int \frac{d^{2}{\bf k}}{(2\pi)^{2}}  \mathcal{A}^{a}_1\mathcal{A}^{b}_1 s^{\beta(-k^2)} \left[i +  \tan\left(\frac{\pi}{2} \beta(-k^2)\right)\right]\right\} \equiv e^{2 i \delta(s,b)}\,\,\,.
\eeqn
Let's now come to the AGK cutting rules.
Since $\Gamma_{\{\beta_{i}\}}$ is real, the full imaginary part of (\ref{obt}) is simply given by replacing $\zeta$ by 1. The AGK rules tell us how this full imaginary part is built up by individual contribution in which the imaginary part of a subset of Reggeon signature factors is taken (i.e. in which a subset of $k$  gravi-reggeons is ``cut" out of the total number $n$).

\be{cgr}
{\rm disc}_{s}[\mathcal{A}^{ab}_{n}(s,t)]=\sum_{k=0}^{n}a_{k}^{n}(s,t)\,\,\,,
\ee
where 
\be{aa}
a_{k}^{n}=2\pi i \int d \Omega_{n} s^{2+\omega} (F_{AGK})_{k}^{n}\mathcal{A}_{n}^{a}\mathcal{A}_{n}^{b}\,\,\, ,
\ee
with
\be{AGK}
(F_{AGK})_{k}^{n}=\frac{2^{n}}{n!}(-1)^{n}+{1\over {n!}}\Gamma_{\{\beta_j\}} \quad {\rm for} \quad k=0\,\,\,,
\ee
$$(F_{AGK})_{k}^{n}=(-1)^{n-k}\frac{2^{n}}{(n-k)!k!} \quad {\rm for} \quad n\ge k>0\,\,\,.$$
Clearly the sum in (\ref{cgr}) reproduces the total discontinuity.

Also for the AGK rules we can go over to impact parameter.
Under the eikonal approximation leading to (\ref{FTexp}) we get \footnote{ A more intuitive  direct derivation of (\ref{AGKb})  was given already in Sect. 2.2 starting from a more detailed cutting formula.},
\be{AGKb}
(S S^{\dagger})_{k}^{n}=(-1)^{n-k}\frac{(4{\rm  Im} \delta) ^{n}}{(n-k)!k!} \quad {\rm for} \quad n\ge k>0\,\,\,.
\ee
Keeping $k$ fixed and summing over $n \ge k$ we get:
\be{sigmak}
(S S^{\dagger})_{k} = e^{-4{\rm  Im} \delta} \frac{(4{\rm  Im} \delta) ^{k}}{k!} \,\,\,.
\ee
Finally, summing over $k \ge 1$ and adding the elastic cross section $\sigma_{el} \equiv  |S_{el}|^2 =  e^{-4{\rm  Im} \delta} $ we recover (s-channel) partial wave unitarity.

Expression (\ref{sigmak}) is interpreted as the probability of having $k$ cut gravi-Reggeons 
at impact parameter ${\bf b}$. 
A Poiss\'on distribution of CGR is obtained.
The mean and the variance are given by
\be{mean}
\langle k \rangle (s,{\bf b})={\rm Var}[k(s,{\bf b})]=4{\rm Im}\delta\,\,\,.
\ee

\section{Building the $B_{0}$ factor}

In this appendix, we will derive the $B_{0}$-factor in the general case of massive particles, take the  massless limit and then show that it is extremized in particular kinematical regimes. 

\subsection{Derivation of $B_0$ for massive and massless particles}

For massless particles, the individual integrals, for fixed $i$ and $j$, are logarithmically divergent. They are given by
\be{Bintegral}
I_{\rm m-less}(p_i, p_j) = \int{d^3q\over |q| qp_i qp_j } = \int {d^3q\over |q|^3 E_iE_j(1-{{\vec n}}\vec{n}_i)(1-{{\vec n}}\vec{n}_j) } \ee
$$= {1\over E_iE_j} \int_\lambda^\Lambda {d|q| \over |q|} \int {d\Omega_\vec{n} \over(1-{{\vec n}}\vec{n}_i)(1-{{\vec n}}\vec{n}_j)}\,\,\,,$$
where $\lambda$ is an IR regulator and $\Lambda$ is the upper limit for the validity of the leading soft behaviour. The angular integral is also logarithmically divergent and reads
$$
J_{\rm m-less}(p_i, p_j) = \int {d\Omega_\vec{n} \over(1-{{\vec n}}\vec{n}_i)(1-{{\vec n}}\vec{n}_j)} = \int_0^1 d\alpha \int {d\cos\theta d\phi \over [1 - \cos\theta |\alpha \vec{n}_i + (1-\alpha) \vec{n}_j|]^2}  = $$ $$ =\int_0^1 { 2\pi d\alpha \over |\alpha \vec{n}_i + (1-\alpha) \vec{n}_j|}  \left[{1\over 1 - |\alpha \vec{n}_i + (1-\alpha) \vec{n}_j|} - {1\over 1 + |\alpha \vec{n}_i + (1-\alpha) \vec{n}_j|} \right] = 
 $$
 $$=  \int_0^1 { 4\pi d\alpha \over 1 - |\alpha \vec{n}_i + (1-\alpha) \vec{n}_j|^2} = 
 \int_0^1 { 2\pi d\alpha \over \alpha(1 -\alpha) (1- \vec{n}_i\vec{n}_j)} = {2\pi  \log \mu \over  (1- \vec{n}_i\vec{n}_j) }\,\,\,.$$ 
 
In order to regulate the logarithmic divergence of the angular integral, it is convenient to treat the `hard' particles as massive and later on take the massless limit. In practice the only difference is that $\vec{n}_i$ is replaced by $\vec{v}_i = \vec{p}_i/E_i$ with $ |\vec{v}_i|<1$
 $$
J_{\rm m-ive}(p_i, p_j) = \int {d\Omega_\vec{n} \over(1-{{\vec n}}\vec{v}_i)(1-{{\vec n}}\vec{v}_j)} = \int_0^1 d\alpha \int {d\cos\theta d\phi \over [1 - \cos\theta |\alpha \vec{v}_i + (1-\alpha) \vec{v}_j|]^2}  = $$ $$ =\int_0^1 { 2\pi d\alpha \over |\alpha \vec{v}_i + (1-\alpha) \vec{v}_j|}  \left[{1\over 1 - |\alpha \vec{v}_i + (1-\alpha) \vec{v}_j|} - {1\over 1 + |\alpha \vec{v}_i + (1-\alpha) \vec{v}_j|} \right] = 
 $$
 $$=  \int_0^1 { 4\pi d\alpha \over 1 - |\alpha \vec{v}_i + (1-\alpha) \vec{v}_j|^2} = 
 - \int_0^1 { 4\pi d\alpha \over \alpha^2|\vec{v}_i-\vec{v}_j|^2 + 2 \alpha(\vec{v}_i-\vec{v}_j)\vec{v}_j + 
|\vec{v}_j|^2 - 1} = $$ 
$$ = { 4\pi \over |\vec{v}_i-\vec{v}_j|^2} { 1 \over \alpha_+ - \alpha_-} \log{ (1- \alpha_-) \alpha_+ \over 
(1- \alpha_+) \alpha_-}  = { 2\pi E_i E_j \over \beta_{ij}} \log{ 1+ \beta_{ij}\over 
1- \beta_{ij}} \,\,\,,$$ 
 where $$|\vec{v}_i-\vec{v}_j|^2 \alpha_{\pm} = (\vec{v}_i-\vec{v}_j)\vec{v}_j \pm \sqrt{ [(\vec{v}_i-\vec{v}_j)\vec{v}_j]^2 + (1- |\vec{v}_j|^2)|\vec{v}_i-\vec{v}_j|^2}\,\,\,,$$
 with
 $$
 \alpha_+\alpha_- = {|\vec{v}_j|^2 - 1 \over |\vec{v}_i-\vec{v}_j|^2}
 \qquad , \qquad 
{\alpha_+ + \alpha_- \over 2} = - {(\vec{v}_i-\vec{v}_j)\vec{v}_j \over |\vec{v}_i-\vec{v}_j|^2} \: ,
$$
 $$
 {\alpha_+ - \alpha_- \over 2} = {\sqrt{ |\vec{v}_i-\vec{v}_j|^2 + (\vec{v}_i\vec{v}_j)^2
 -|\vec{v}_i|^2|\vec{v}_j|^2}\over |\vec{v}_i-\vec{v}_j|^2} =   {\beta_{ij} (1- \vec{v}_i\vec{v}_j) \over |\vec{v}_i-\vec{v}_j|^2}\,\,\,,
 $$
 since 
 $$
 \beta_{ij}^2 = 1 - {m_i^2 m_j^2 \over (p_ip_j)^2} = 1 - {(1-|\vec{v}_i|^2)(1-|\vec{v}_j|^2) \over 
 (1-\vec{v}_i\vec{v}_j)^2} = {|\vec{v}_i-\vec{v}_j|^2 + (\vec{v}_i\vec{v}_j)^2 -|\vec{v}_i|^2|\vec{v}_j|^2 \over 
 (1-\vec{v}_i\vec{v}_j)^2}
\,\,\,. $$
Moreover one finds 
 $$
 { (1- \alpha_-) \alpha_+ \over (1- \alpha_+) \alpha_-}  
 = { {1\over 2}(\alpha_+ + \alpha_-) + {1\over 2}(\alpha_+ - \alpha_-) - \alpha_-\alpha_+ \over {1\over 2}(\alpha_+ + \alpha_-) - {1\over 2}(\alpha_+ - \alpha_-) - \alpha_-\alpha_+} 
 =  
{ {1-\vec{v}_i\vec{v}_j \over |\vec{v}_i-\vec{v}_j|^2} + {(1-\vec{v}_i\vec{v}_j) \beta_{ij} \over |\vec{v}_i-\vec{v}_j|^2} \over {1-\vec{v}_i\vec{v}_j \over |\vec{v}_i-\vec{v}_j|^2} - {(1-\vec{v}_i\vec{v}_j) \beta_{ij} \over |\vec{v}_i-\vec{v}_j|^2} } = {1 + \beta_{ij} \over 1- \beta_{ij}}\,\,\,.
$$
Including the overall ($q$-independent and thus unintegrated) factors 
$${8\pi G \eta_{i}\eta_{j}[(p_{i}{\cdot} p_{j})^{2}-\frac{1}{2}m_{i}^{2}m_{j}^{2}]\over 2(2\pi)^{3} E_i E_j} =
{G \eta_{i}\eta_{j}m_{i}m_{j}(1+\beta_{ij}^{2}) \over (2\pi)^2 (1-\beta_{ij}^{2})^{1/2}}
$$ 
and summing over $i$ and $j$ lead to Weinberg's celebrated result (\ref{onecon}) \cite{Weinberg:1965nx} and to its massless limit, used in the present investigation:
\beqn{oneconappendix}
B_{0} = \frac{G}{2\pi}\sum_{i,j} \eta_{i}\eta_{j}m_{i}m_{j}\frac{1+\beta_{ij}^{2}}{\beta_{ij}(1-\beta_{ij}^{2})^{1/2}}\log \left(\frac{1+\beta_{ij}}{1-\beta_{ij}}\right) \: \rightarrow \: 
\frac{2G}{\pi}\sum_{i,j} \eta_{i}\eta_{j}p_{i}p_{j}\log{p_{i}p_{j} \over \mu^2}.\eeqn

\subsection{Special kinematics for massless particles}

Analyzing the first order constrained variation of $B_0$ in the massless case 
around collinear and orthogonal kinematical configurations, we here show that they extremize $B_0$.

First of all, let us rewrite the  Eq. (\ref{conv}) as
\beqn{B0N}
B_{0} &=&  {2G s \over\pi}  \left\{ {-} \sum_{{i}} w_{{i}} \left[ (1{-}\vec{n}\vec{n}_{{i}}) \log w_{{i}} (1{-}\vec{n}\vec{n}_{{i}}) + (1{+}\vec{n}\vec{n}_{{i}}) \log w_{{i}} 
(1{+}\vec{n}\vec{n}_{{i}})  \right] \right. \nonumber \\
&+& \left. {1\over 2} \sum_{{i},{j}} 2 w_{{i}} w_{{j}}(1{-}\vec{n}_{{i}}\vec{n}_{{j}} )  \log 2 w_{{i}} w_{{j}}(1{-}\vec{n}_{{i}}\vec{n}_{{j}} ) \right\}\,\,\,.
\eeqn
{\it }
In fact, using momentum conservation, one can further simplify the above expression and arrive at  
\beqn{B0senzaw}
B_{0} &=&  {2G s \over \pi}  \left\{ {-} \sum_{{i}} w_{{i}} \left[ (1{-}\vec{n}\vec{n}_{{i}}) \log {1{-}\vec{n}\vec{n}_{{i}}\over 2} + (1{+}\vec{n}\vec{n}_{{i}}) \log {1{+}\vec{n}\vec{n}_{{i}}\over 2}  \right] \right. \nonumber \\
&+& \left. \sum_{{i},{j}} w_{{i}} w_{{j}}(1{-}\vec{n}_{{i}}\vec{n}_{{j}} )  \log{1{-}\vec{n}_{{i}}\vec{n}_{{j}}\over 2} \right\}\,\,\,,
\eeqn
which can be more compactly written as in Eq. (\ref{conv})

Let us then consider the first order constrained variation of the $B_{0}$-factor.
In general, keeping $N$ as well as $p_1$ and $p_2$ fixed and varying the momenta of the outgoing particles, one has $p_i \rightarrow p_i +\delta p_i$ with $p_i{\cdot}\delta p_i = 0$ ($p_i^2 =0$) and $\sum_i\delta p_i = 0$.
Setting  $\delta p_i = (u_i, \vec{v}_i) = u_i (1, \vec{n}_i) + (0, \vec{q}_i) = \delta p^\parallel_i + \delta p^\perp_i$  with $u_i = \vec{v}_i\vec{n}_i$ and $ \vec{q}_i = \vec{v}^\perp_i = \vec{v}_i - u_i \vec{n}_i$ so that  $\vec{q}_i\vec{n}_i =0 $. 

The (constrained) first order variation of $B_{0}$, 
$$\delta B_0 = \left. \sum_i {\partial B_0 \over \partial p_i} \delta p_i\right\vert_{\rm constrained} $$ $$= {2G \over\pi} \left\{ 
- \sum_i [p_{1} \delta p_i (\log p_{1}p_i  +1)+ p_{2} \delta p_i (\log p_{2}p_i + 1) ]+
{1\over 2}\sum_{i,j} (p_{j} \delta p_i + p_{i} \delta p_j)(\log p_{i}p_j + 1)\right\}\,\,\,,
$$
the terms with 1 drop thanks to momentum conservation $\sum_i \delta p_i =0$, so that 
$$\delta B_0 = {2G \over \pi} \left\{ 
- \sum_i [p_{1} \delta p_i \log p_{1}p_i + p_{2} \delta p_i \log p_{2}p_i ]+
{1\over 2}\sum_{i,j}(p_{j} \delta p_i + p_{i} \delta p_j)\log p_{i}p_j \right\}\,\,\,.
$$

\vspace{0.5cm}
{\bf Near-collinear kinematics}

Let us further specialise to the case of a near-collinear kinematical scattering.
For a collinear configuration $\vec{n}_i = \sigma_i \vec{n}$. 
As a result, $\delta p^\perp_i$ does not contribute and one has $p_{1,2} \delta p_i = p_{1,2} \delta p^\parallel_i = E u_i (1-\sigma_i)$ and $p_{j} \delta p_i = p_{j} \delta p^\parallel_i = E_j u_i (1-\sigma_j\sigma_i)$. Plugging into $\delta B_0$ one finds a vanishing result
$$\delta B_0 = {2G E \over \pi} \left\{ 
- \sum_i u_i [(1-\sigma_i)\log 2w_i {1{-}\sigma_i\over 2} + (1+\sigma_i)\log 2w_i {1{+}\sigma_i\over 2}] \right.
$$ 
$$\left. +\sum_i (w_{j} u_i + w_{i} u_j)(1-\sigma_i \sigma_j) \log 4w_{i}w_{j}{1{-}\sigma_i\sigma_j\over 2}\right\} = 0 \, ,
$$
since the constant terms with $\log 2$ vanish due to the constraints $\sum_i u_i = 0 = \sum_i u_i \sigma_i$, the terms with $u_i\log w_i$ cancel each other and the terms $(1\pm \sigma) \log (1\pm\sigma / 2) = 0$ for $\sigma = \pm 1$. This proves that collinear configurations extremize $B_0$. 
The second order variations along the longitudinal directions yield a vanishing result: the particles remain collinear and any choice of $w_i$ and $\sigma_i$ such that $\sum_i w_i =1$ and $\sum_i w_i\sigma_i=0$ produces $B_0 =0$. Variations of collinear configurations along $\delta^\perp p_i$ produce a positive result: $B^{\rm coll}_0 =0$ is a local infinitely degenerate minimum and $B^{\rm near-coll}_0>0$.

{\bf Near-orthogonal kinematics}

Let us consider the case of a near-orthogonal kinematical scattering.
For a configuration with two back-to-back jets perpendicular to the direction of the incoming particles $\vec{n}_i = \sigma_i \vec{m}$ with $\vec{n}\vec{m} = 0$. 
As before $\delta p_i = \delta p^\parallel_i + \delta p^\perp_i = $ so that $p_{j} \delta p_i = p_{j} \delta p^\parallel_i = E_j u_i (1-\sigma_j\sigma_i)$ while $p_{1} \delta p_i = E (u_i - \vec{m}\vec{q}_i)$ and $p_{1} \delta p_i = E (u_i + \vec{m}\vec{q}_i)$. The (constrained) first order variation of $B_{0}$, 
$$\delta B_0 = {2G \over \pi} \left\{ 
- \sum_i E \log w_i [(u_i {-} \vec{m}\vec{q}_i) + (u_i {+} \vec{m}\vec{q}_i) \log w_i]  \right.+$$ $$ \left. 
\sum_{i,j} (E_j u_i + E_i u_j) {1-\sigma_j\sigma_i \over 2} \log 4 w_{i} w_j {1-\sigma_j\sigma_i \over 2}\right\} = {2G \over \pi} \left\{ 
- \sum_i 2 E u_i \log w_i + \right.
$$
$$
\left. +\sum_{i,j} (E_j u_i + E_i u_j) {1-\sigma_j\sigma_i \over 2} [\log 4 + \log w_{i} + \log w_j + \log {1-\sigma_j\sigma_i \over 2}\right\} =0 \, ,
$$
since the constant terms with $\log 4$ vanish due to the constraints $\sum_i u_i = 0 = \sum_i u_i \sigma_i$, the terms with $u_i\log w_i$ cancel each other and the terms $(1\pm \sigma) \log (1\pm\sigma / 2) = 0$ for $\sigma = \pm 1$. This proves that `orthogonal' configurations extremize $B_0$ but are highly degenerate. 
The second order variations of the outing particles along the direction of the two jets (orthogonal to the beam-line) yield a vanishing result:  and any choice of $w_i$ and $\sigma_i$ such that $\sum_i w_i =1$ and $\sum_i w_i\sigma_i=0$ produces $B_0 = {2Gs \over\pi} \log 2$. This is to be expected, since the two jets of collinear particles behave as two particles with energy $E$ and momentum $\vec{p}$ orthogonal to $\vec{p}_{1,2}$.

\end{document}